\documentclass[prd,twocolumn,eqsecnum,nofootinbib]{revtex4}
\usepackage{bm} 
\usepackage{amssymb}
\usepackage{graphicx} 
\newcommand{\unit}[1]{\bm{\hat{#1}}}
\newcommand{\p}{{\sf p}} 
\renewcommand{\t}{{\sf t}} 
\begin{document}
\title{Multi-scale analysis of the electromagnetic self-force in
  a weak gravitational field} 
\author{Adam Pound and Eric Poisson}
\affiliation{Department of Physics, University of Guelph, Guelph, 
Ontario, Canada N1G 2W1}
\date{December 18, 2007} 
\begin{abstract} 
We examine the motion of a charged particle in a weak gravitational 
field. In addition to the Newtonian gravity exerted by a large central 
body, the particle is subjected to an electromagnetic self-force that
contains both a conservative piece and a radiation-reaction
piece. This toy problem shares many of the features of the
strong-field gravitational self-force problem, and it is sufficiently
simple that it can be solved exactly with numerical methods, and
approximately with analytical methods. We submit the equations of
motion to a multi-scale analysis, and we examine the roles of the
conservative and radiation-reaction pieces of the self-force. We show
that the radiation-reaction force drives secular changes in the
orbit's semilatus rectum and eccentricity, while the conservative
force drives a secular regression of the periapsis and affects the
orbital time function; neglect of the conservative term can hence give
rise to an important phasing error. We next examine what might be
required in the formulation of a reliable secular approximation for
the orbital evolution; this would capture all secular changes in the
orbit and discard all irrelevant oscillations. We conclude that such
an approximation would be very difficult to formulate without prior 
knowledge of the exact solution.  
\end{abstract} 
\maketitle

\section{Introduction} 

The gravitational inspiral of a solar-mass compact object into a
massive black hole residing in a galactic center has been identified
as one of the most promising sources of gravitational waves for the
Laser Interferometer Space Antenna \cite{LISA}. The need for accurate  
theoretical models of the expected signal, for the purposes of signal 
detection and source identification, has motivated an intense effort
from many workers to determine the motion of the small body in the
field of the large black hole. This is done in a treatment that goes
beyond the geodesic approximation and takes into account the body's
own gravitational field, which is a small perturbation over the field
of the black hole. In this treatment the small body can be described
as moving on an accelerated world line in the background spacetime of
the large black hole; the body is said to move under the influence of
its own gravitational self-force \cite{mino-etal:97,
  quinn-wald:97}, and this force is derived from the retarded
gravitational perturbation produced by the moving body. For a review 
of the self-force formalism, see Ref.~\cite{poisson:04b} and the
special issue of {\it Classical and Quantum Gravity} devoted to this
topic \cite{lousto:05}.     

The concrete evaluation of the gravitational self-force acting on a
small body moving in the Kerr spacetime is a challenging project that
has not yet been completed (although progress has been steady). Given
the severity of the challenge, a number of authors \cite{mino:03,
  mino:05a, mino:05b, drasco-etal:05, hughes-etal:05, sago-etal:05, 
  drasco-hughes:06, drasco:06, sago-etal:06, ganz-etal:07,
  babak-etal:07} 
have attempted to formulate various simple schemes that would allow
them to reproduce the effects of the self-force; their hope is that
these schemes will be simple enough for rapid implementation in
numerical codes, and accurate enough to describe faithfully the
orbital evolution of a body subjected to its gravitational
self-force. One such scheme is {\it Mino's radiative approximation}
\cite{mino:03, mino:05a, mino:05b}, which is based on an approximate
self-force constructed from the half-retarded minus half-advanced
gravitational perturbation associated with the moving body. Mino
proved that while his version of the self-force neglects all
conservative corrections to the motion, it correctly accounts for the
long-term dissipative effects associated with the true
self-force. This led to the widespread belief that all long-term
secular changes in the orbital motion would be captured by the
radiative approximation, and that conservative effects would produce
only short-term changes that would not accumulate in the long run. The
simplicity of Mino's scheme made it attractive, and it was adopted in
a number of works that aimed to model the inspiral of a small body
into a rapidly rotating black hole \cite{drasco-etal:05,
  hughes-etal:05, sago-etal:05, drasco-hughes:06, drasco:06,
  sago-etal:06, ganz-etal:07, babak-etal:07}. 

Mino's radiative approximation was criticized, however, in an earlier
work by the present authors \cite{pound-poisson-nickel:05}, hereafter
referred to as ``paper I.'' Building on an analogy between the
gravitational self-force and its electromagnetic counterpart, we
showed that conservative terms in the true self-force do lead to
long-term secular changes in the orbital motion. These changes are not
captured by the radiative approximation, and we concluded that Mino's
scheme has severe limitations. This conclusion was supported by a
recent analysis by Drasco and Hughes \cite{drasco-hughes:06}, and the
general attitude currently is that while the radiative approximation
may be useful to generate template waveforms for signal detection, it
is probably insufficient for reliable parameter estimation.  

Our purpose in this paper is to revisit the analysis presented in 
paper I. There are three reasons for reopening the case. The first is
that our original analysis of the electromagnetic self-force employed
rather crude mathematical tools, and we wish to present here a more
thorough and rigorous treatment. The second is that the main source of 
discrepancy between the effects of the radiative self-force and those
of the true self-force was not correctly identified in paper I. In the
original paper we claim that the discrepancy is mainly due to the
secular regression in the orbit's periapsis, an effect that is 
produced by the true self-force but not accounted for by the radiative
approximation. In this paper we show that while this is indeed a
source of discrepancy, it is not the most important one. As we shall 
explain in Sec.~IV C, the most important conservative effect is
actually associated with the time function on the orbit. The third
reason is that we wish to introduce here a clear distinction between
the {\it radiative approximation} to the self-force and the notion of
a {\it secular approximation} to an orbital evolution; our secular
approximation is a specific implementation of the general idea of
capturing the long-term orbital evolution through an {\it adiabatic
approximation} that allows the orbit to evolve slowly. The phrases  
``radiative approximation'' and ``adiabatic approximation'' are used
synonymously in paper I (and indeed, in most of the literature on this
topic), but we feel that this is a highly misleading practice. A large
portion of this paper is devoted to the task of identifying what
should be required of a good secular approximation, and we shall see
that the radiative approximation does not meet those requirements.  

The precise meaning of an adiabatic approximation is somewhat
ambiguous in the literature. In all cases, the basic assumption is
that the secular effects of the self-force occur on a time scale that
is long compared to the orbital period. From this assumption, numerous
approximations have been formulated: (1) Since the particle's orbit
deviates only slowly from geodesic motion, the self-force can be
calculated as if the particle travels on a geodesic (or, in the
post-Newtonian case, the radiation reaction can be calculated as if
the particle's dynamics were conservative); (2) since the
radiation-reaction time scale is much longer than the orbital period,
periodic effects can be neglected; and (3), based on various
arguments, conservative effects can be neglected. Although each one of
these three approximations has been called an adiabatic approximation,
we believe that they should be distinguished from one another. To
discuss the first approximation is beyond the scope of this paper. We
focus instead on the latter two approximations: number (2) above,
which we call the \textit{secular approximation}, and which neglects
periodic effects; and number (3), which we shall call the 
\textit{radiative approximation}, and which neglects conservative
effects.   

The main idea behind the construction of a secular approximation is
the following. We consider an orbital evolution under the action of a 
self-force, and we wish to simplify the equations of motion in such a
way that the long-term, secular changes will be captured, at the cost 
of discarding irrelevant, short-term effects. Suppose that we describe
the orbital evolution in terms of a set of orbital elements $I^A(t)$,
where $A$ is an index that labels each element. (This description is
introduced in Sec.~III, and explained fully in Appendices A and B.)
The orbital elements would be constant in the absence of a perturbing
force, but they evolve in time as a result of the force's action. It
is expected that each orbital element will display a behavior that can
be decomposed into a secular change that accumulates monotonically
over time, and an oscillation that averages to zero in the long
run. We thus write $I^A(t) = I^A_{\rm sec}(t) + I^A_{\rm osc}(t)$, and
a secular approximation for the orbital elements would keep the
secular terms and discard the oscillations. We would write 
$I^A(t) \simeq I^A_{\rm sec}(t)$, and seek a method to obtain  
$I^A_{\rm sec}(t)$ in the most direct and convenient way
possible. Presuming that this must be done in a context in which the 
exact solution $I^A(t)$ would be too difficult to obtain, we would
seek to formulate equations of motion directly for $I^A_{\rm sec}(t)$, 
and we would hope that those equations are sufficiently simple that a    
solution could easily be found (analytically or numerically). This is
the main idea, and the task ahead appears to be clearly
identified. But to turn the idea into a precise algorithm may not be
easy. To illustrate the difficulty we shall examine, in a specific
context in which we can make progress (the electromagnetic self-force
of paper I), what would be required in the construction of a secular
approximation for the orbital evolution.   

The secular approximation is logically distinct from the radiative
approximation, in which the true self-force is truncated so as to
discard all conservative terms. In the radiative approximation, one
writes $I^A(t) \simeq I^A_{\rm rad}(t)$, and one calculates 
$I^A_{\rm rad}(t)$ on the basis of the truncated self-force. It is
known, as Mino has shown \cite{mino:03, mino:05a, mino:05b}, that the 
radiative self-force correctly accounts for the long-term, 
{\it dissipative changes} in the orbital elements. If it correctly
produced the long-term, {\it conservative changes} as well, we
would conclude that the radiative approximation captures the idea of
a secular approximation. But, as we have shown in paper I
\cite{pound-poisson-nickel:05}, and as we intend to show even more
convincingly here, the radiative approximation fails to account for
secular changes in $I^A(t)$ that are produced by the conservative
piece of the self-force. The radiative and secular approximations
are therefore distinct, and we consider it important to distinguish
these terms carefully.    

We will examine the limitations of the radiative approximation, and
attempt to construct a faithful secular approximation, in the
specific context of an electromagnetic self-force acting on a charged   
particle moving in a weak gravitational field. The motion of the
particle is governed by the equations
\begin{equation} 
\bm{a} = \bm{g} + \bm{f}_{\rm self}, 
\label{1.1}
\end{equation}
where $\bm{a} = d^2\bm{r}/dt^2$ is the particle's acceleration vector, 
$\bm{g} = -M\bm{\hat{r}}/r^2$ is the Newtonian gravitational field of
a body of mass $M$, and 
\begin{equation} 
\bm{f}_{\rm self} = \lambda_{\rm c} \frac{q^2}{\mu} 
\frac{M}{r^3} \bm{\hat{r}} 
+ \lambda_{\rm rr} \frac{2}{3} \frac{q^2}{\mu} \frac{d \bm{g}}{dt}  
\label{1.2}
\end{equation}
is the electromagnetic self-force divided by the particle's mass
$\mu$. (The particle is treated as a test mass. In this treatment, the
central body does not move, its mass $M$ is the system's total mass,
and $\mu$ is the system's reduced mass.) This self-force was
calculated \cite{dewitt-dewitt:64, pfenning-poisson:02} for the weakly
curved spacetime produced by the central body, assuming that the
charged particle moves slowly. Here $q$ is the particle's charge, and
$\bm{r}(t)$ is its position vector relative to the central body; we
have also introduced the distance $r = |\bm{r}|$ and the unit vector
$\bm{\hat{r}} = \bm{r}/r$. The constants $\lambda_{\rm c}$ and
$\lambda_{\rm rr}$ in Eq.~(\ref{1.2}) are both equal to unity; they
serve to remind us that the first term in Eq.~(\ref{1.2}) is the
conservative piece of the self-force, while the second term is the
dissipative (or radiation-reaction) piece. By keeping these constants
in our calculations we will be able to distinguish conservative
effects from dissipative effects; the radiative approximation is
obtained by setting $\lambda_{\rm c} = 0$ and keeping 
$\lambda_{\rm rr} = 1$. Throughout the paper we use the usual
vectorial notation of three-dimensional flat space, and we work in
units such that $G = c = 1$.   

We work in the specific context of Eqs.~(\ref{1.1}) and (\ref{1.2})
for two reasons. First, our toy problem is an actual example of a 
self-force acting on an orbiting body. While the force has an
electromagnetic origin instead of a gravitational origin, and while
the motion takes place in a weak (Newtonian) gravitational field
instead of a strong field, the self-force of Eq.~(\ref{1.2})
nevertheless contains conservative and dissipative terms that will
have different effects on the orbital motion. In the usual Lorenz
gauge, the gravitational self-force in strong fields will also contain
conservative and dissipative pieces. In addition, each self-force
comes with a similar post-Newtonian counting. From Eq.~(\ref{1.2}) we
see that the conservative term in the electromagnetic self-force is a
correction of order $q^2/(\mu r)$ relative to $\bm{g}$, and taking $q$
to be of order $\mu$, we recognize this as a correction of first
post-Newtonian (1{\sc pn}) order; the dissipative term is a correction
of order $q^2 v/(\mu r)$, where $v$ is the orbital velocity, and we
recognize this as a correction of 1.5{\sc pn} order. On the other
hand, in a post-Newtonian context the gravitational self-force
presents conservative pieces at orders 0{\sc pn}, 1{\sc pn}, 
2{\sc pn}, 3{\sc pn}, and so on, and dissipative pieces at orders
2.5{\sc pn}, 3.5{\sc pn}, and so on. While the post-Newtonian counting
is not identical, in each case we have dominance of the conservative
effects over the dissipative effects, and all in all, this gives us
good reasons to believe that the electromagnetic problem captures 
the essential physics of the more complicated, gravitational
problem when it is formulated in the usual Lorenz gauge.  

Second, Eq.~(\ref{1.1}) is far simpler than the realistic
self-force equation (for which the force must be obtained
numerically), and this permits a very thorough and rigorous 
mathematical analysis. We shall therefore be able to extract very
precise consequences of Eqs.~(\ref{1.1}) and (\ref{1.2}), and examine
closely the issues that concern us regarding the radiative and
secular approximations. The simple mathematics of the toy problem
will allow us to draw firm and clear conclusions, and the proximity of
its physics to that of the realistic problem will give us confidence
that these conclusions extend from the toy problem to the realistic
case.  

We begin in Sec.~II with a simple illustration of the themes to be 
explored in this paper. The mathematical analysis of Eqs.~(\ref{1.1})
and (\ref{1.2}) is carried out in Sec.~III, in the framework of
osculating orbital elements summarized in Appendices A and B. The
mathematical details are presented in Sec.~III with minimum
commentary, but we present a full discussion of our results in
Sec.~IV. We summarize our conclusions in Sec.~V.   

\section{Illustration} 

Before we proceed with our mathematical analysis of Eqs.~(\ref{1.1})
and (\ref{1.2}), it is helpful to describe a very simple problem that
illustrates rather well the issues we shall encounter.   

Suppose that we are interested in a quantity $q(t)$ that is governed
by the system of dynamical equations 
\begin{equation} 
\frac{dq}{d\lambda} = \epsilon_1 - \epsilon_2 \sin\lambda, \qquad 
\frac{dt}{d\lambda} = 1 + \cos\lambda + \epsilon_3, 
\label{2.1}
\end{equation}
where $\lambda$ is a running parameter, and $\epsilon_1$,
$\epsilon_2$, and $\epsilon_3$ are small constants. The differential 
equations come with the initial conditions 
\begin{equation} 
q(0) = 1, \qquad t(0) = 0. 
\label{2.2}
\end{equation} 
In this example, $q(t)$ is analogous to the set of orbital elements
$I^A(t)$ that were introduced previously, and the equation for
$dq/d\lambda$ is analogous to Eq.~(\ref{3.10}) below, with 
$\epsilon_1$ and $\epsilon_2$ playing the roles of $\epsilon_{\rm rr}$ 
and $\epsilon_{\rm c}$, respectively. The quantity $q$ is constant
when $\epsilon_1 = \epsilon_2 = 0$ (the unperturbed situation), and it
acquires a time dependence when the perturbation is turned on. The
equation for $dt/d\lambda$ is analogous to the first line of
Eq.~(\ref{3.34}) below, with $\epsilon_3$ playing the role of
$\epsilon_{\rm c}$. As we explain in Sec.~III C, the first term
proportional to $\epsilon_{\rm c}$ in Eq.~(\ref{3.34}) is generated by 
oscillatory terms in the orbital elements and oscillatory terms in the
unperturbed equation for $t$; these combine to give rise to a secular
term in the perturbed equation. To ignore the oscillations would
produce the significant mistake of dropping the $\epsilon_3$ term in 
Eq.~(\ref{2.1}). The variable $\lambda$ gives a convenient
parameterization of the motion; its use is motivated by the fact that
the ``force'' can be expressed as a simple function of $\lambda$,
while it would be very difficult to express it in terms of the time
variable $t$. The parameter $\lambda$ is analogous to the orbital
parameter $\phi$ that will be introduced in Sec.~III. (It is also
analogous to ``Mino time'' \cite{mino:03, drasco-hughes:04}, a
convenient parameterization of geodesics in Kerr spacetime.) 

The exact solution to the system of equations is 
\begin{eqnarray}
q(\lambda) &=& 1 + \epsilon_1 \lambda + \epsilon_2(\cos\lambda - 1),
\label{2.3} \\  
t(\lambda) &=& \lambda + \sin\lambda + \epsilon_3 \lambda. 
\label{2.4}
\end{eqnarray} 
We see that $\epsilon_1$ produces a secular growth in $q(\lambda)$,
$\epsilon_2$ is associated with oscillations, and $\epsilon_3$
produces a secular drift in the time function $t(\lambda)$.  

Suppose that we are interested only in the long-term, secular changes
in $q$, and that we wish to construct a secular approximation for
it. Because we have access to the exact solution, it is a simple
matter to remove the oscillations by subjecting it to an averaging
procedure. Our first option is to introduce  
\begin{equation}
\langle q \rangle_\lambda := \frac{1}{2\pi} 
\int_{\lambda-\pi}^{\lambda+\pi} q(\lambda')\, d\lambda' 
\label{2.5}
\end{equation} 
and to define the secular approximation as 
$q_{\rm sec} = \langle q \rangle_\lambda$. Our choice here is
therefore to remove the oscillations with respect to $\lambda$, and a
calculation based on Eqs.~(\ref{2.3}) and (\ref{2.5}) gives 
\begin{equation} 
\langle q \rangle_\lambda = 1 + \epsilon_1 \lambda - \epsilon_2 
+ O(\epsilon^2). 
\label{2.6}
\end{equation} 
This version of the secular approximation is a solution to the
modified differential equation  
\begin{equation} 
\frac{d}{d\lambda} \langle q \rangle_\lambda = \epsilon_1 
+ O(\epsilon^2) 
\label{2.7}
\end{equation}
with the modified initial condition 
\begin{equation} 
\langle q \rangle_\lambda(0) = 1 - \epsilon_2 + O(\epsilon^2). 
\label{2.8}
\end{equation} 
If we did not have access to the exact solution, we might still have
guessed that the correct differential equation for $q_{\rm sec}$ is 
Eq.~(\ref{2.7}), because it can be obtained directly from 
Eq.~(\ref{2.1}) by averaging over the oscillatory term. But we would 
be hard pressed to guess that the correct initial condition is given
by Eq.~(\ref{2.8}). Using the approximate differential equation with
the exact initial condition $q_{\rm sec}(0) = 1$ would produce a
function that is offset by $\epsilon_2$ relative to 
$\langle q \rangle_\lambda$. 

Our first message is that a faithful secular approximation can be 
based on an averaged version of the differential equation, but that it
must come also with a corresponding change of initial condition. To
obtain the approximate differential equation might be easy, but to
identify the correct initial condition is impossible when the exact
solution is unknown. The formulation of a secular approximation
therefore suffers from an ambiguity regarding the correct choice of
initial condition. In this example the consequence of missing the
$\epsilon_2$ term in the initial condition is not severe: The
difference between the solutions $1 + \epsilon_1 \lambda$ and
$1 + \epsilon_1 \lambda - \epsilon_2$ becomes relatively small
as $\lambda$ increases and each solution grows secularly. In other
situations, however, the difference in initial conditions could lead
to more serious discrepancies.   

In Eq.~(\ref{2.5}) we removed the oscillations of the exact solution
by averaging over the parameter $\lambda$. Because the observer might
be more interested in the time-behavior of the function $q$, an
alternative choice is to perform the averaging over $t$ instead of
$\lambda$. And since $t(\lambda)$ contains oscillations, it should be
expected that this alternative method of averaging will lead to a
distinct formulation of the secular approximation. Our second option
is therefore to introduce  
\begin{equation}
\langle q \rangle_t := 
\frac{\int_{\lambda-\pi}^{\lambda+\pi} q(\lambda') (dt/d\lambda')\,
  d\lambda'}{\int_{\lambda-\pi}^{\lambda+\pi} (dt/d\lambda')\,
  d\lambda'}
\label{2.9}
\end{equation} 
and to define version 2 of the secular approximation as 
$q_{\rm sec} = \langle q \rangle_t$. A calculation based on
Eqs.~(\ref{2.1}), (\ref{2.3}), and (\ref{2.9}) gives  
\begin{equation} 
\langle q \rangle_t = 1 + \epsilon_1 (\lambda - \sin\lambda) 
- \frac{1}{2} \epsilon_2 + O(\epsilon^2). 
\label{2.10}
\end{equation} 
This is a solution to the modified differential equation 
\begin{equation} 
\frac{d}{d\lambda} \langle q \rangle_t = \epsilon_1(1 + \cos\lambda)  
+ O(\epsilon^2) 
\label{2.11}
\end{equation}
and the modified initial condition 
\begin{equation} 
\langle q \rangle_t(0) = 1 - \frac{1}{2}\epsilon_2 + O(\epsilon^2). 
\label{2.12}
\end{equation} 
Here the situation is more interesting. If we did not have access to
the exact solution, we would never have guessed that the correct
differential equation for the secular approximation is
Eq.~(\ref{2.11}), and we would also never have arrived at
Eq.~(\ref{2.12}).  

Our second message is that this new secular approximation (version  
2, which removes the oscillations in $t$ instead of the oscillations
in $\lambda$) must be based on an approximate differential equation 
and an approximate initial condition that are impossible to identify
without knowing the solution to the exact problem. The ambiguity of
the first method extends from the choice of initial condition to the
specification of the differential equation.    

Our third message is that while the idea of formulating a secular
approximation is clear enough, it is difficult to turn it into a
precise algorithm. To remove the oscillations of an exact solution is
easy enough. But to reformulate the system of differential equations
and initial conditions into a set of approximate equations that would 
achieve the same result is difficult; it might well be impossible in
most cases.   

Our fourth message is concerned with the analogue here of formulating 
a radiative approximation to Eqs.~(\ref{2.1}). This is obtained by
setting $\epsilon_2 = \epsilon_3 = 0$ while leaving $\epsilon_1$
unchanged. This produces the functions 
\begin{equation} 
q_{\rm rad}(\lambda) = 1 + \epsilon_1 \lambda, \qquad 
t_{\rm rad}(\lambda) = \lambda + \sin\lambda. 
\label{2.13}
\end{equation} 
After a long time, when $\lambda \gg 1$, $q_{\rm rad}(\lambda)$
becomes very nearly equal to $q(\lambda)$, and the radiative
approximation is accurate when $q$ is expressed in terms of the
orbital parameter. At late times, however, we have that 
$t_{\rm rad}(\lambda) \simeq \lambda$ while $t(\lambda) 
\simeq (1 + \epsilon_3) \lambda$, and we see that the radiative
approximation produces a shift in the time function that becomes
important when $\lambda$ increases beyond $1/\epsilon_3$; we shall see
that this feature is present also in the context of the
electromagnetic self-force in Sec.~III. When $q$ is expressed as a
function of time, we get that $q_{\rm rad}(t) \simeq 1 
+ \epsilon_1 t$, while the exact solution behaves as 
$q(t) \simeq 1 + \epsilon_1 t /(1 + \epsilon_3)$. The difference is
equal to $\epsilon_1 \epsilon_3 t/(1 + \epsilon_3)$. While this
appears to be small because of the first factor of order $\epsilon^2$,
it is steadily growing because of the additional factor of $t$. The
radiative approximation, therefore, produces a secular drift in the
time function, and a corresponding drift in $q$. 

\section{Multi-scale analysis of the electromagnetic self-force} 

We now proceed with our mathematical analysis of Eqs.~(\ref{1.1}) 
and (\ref{1.2}). This section, unlike all others in this paper, is 
highly technical, and we intend to deal with the technical issues  
while keeping the commentary to a minimum. The implications of our
results, in the light of the themes introduced in Sec.~I, will be
fully detailed in Sec.~IV. The reader who may not wish to delve into
the technical details, and who would prefer to pick up the story where
we left off at the end of Sec.~II, can omit reading this section and
proceed directly to Sec.~IV.  

\subsection{System of equations} 

We wish to integrate the equations of motion (\ref{1.1}) for the
electromagnetic self-force of Eq.~(\ref{1.2}). We shall do so by
employing the method of osculating orbital elements developed in 
Appendix B. 

The starting point of the method is the unperturbed situation 
described by the equations $\bm{a} = \bm{g}$, in which the particle
follows a Keplerian orbit characterized by a number of orbital
elements. (Kepler's problem is reviewed in Appendix A.) The orbital
elements are constants of the Keplerian motion; they are related to
the initial conditions placed on the particle's position and velocity
vectors, but they are defined so as to provide the most useful
information regarding the geometric properties of the orbit. The
elliptical shape of the Keplerian orbit is described by      
\begin{equation} 
r(\phi) = \frac{p}{1 + e\cos(\phi-\omega)} 
\label{3.1}
\end{equation} 
where $r$ is the distance between the particle and the central mass  
$M$, and $\phi$ is the longitude. The orbital elements are the 
semilatus rectum $p$, the eccentricity $e$, and the longitude at
periapsis $\omega$. The elements $p$ and $e$ determine on which
ellipse the particle is moving, and we shall call them the 
{\it principal orbital elements}. The element $\omega$ determines the
particle's initial position on the selected ellipse, and we shall
refer to it as a {\it positional orbital element}. The position of the 
particle as a function of time is determined by integrating  
\begin{equation} 
t' = \sqrt{\frac{p^3}{M}} \frac{1}{[1 + e\cos(\phi-\omega)]^2} 
\label{3.2}
\end{equation} 
for the time function $t(\phi)$; the prime indicates differentiation
with respect to $\phi$. The motion in the orbital plane is then fully
described (in parametric form) by the functions $r(\phi)$ and
$t(\phi)$. It is an important fact that Eq.~(\ref{3.2}) does not admit
a closed-form solution; a convenient way to handle it is by
straightforward numerical integration.     

We next move on to the equations $\bm{a} = \bm{g} 
+ \bm{f}_{\rm self}$ and a description of the perturbed motion. In the 
method of osculating orbital elements, the motion continues to be
described by Eqs.~(\ref{3.1}) and (\ref{3.2}), but the orbital
elements $(p,e,\omega)$ acquire a $\phi$-dependence that accounts for    
the perturbation. Their evolution equations are given by 
Eqs.~(\ref{B.16})--(\ref{B.18}) in Appendix B. They rely on a
decomposition of the self-force according to $\bm{f}_{\rm self} = R 
\unit{r} + S \unit{\phi}$, with $R$ denoting its radial component and
$S$ its tangential component; the unit vectors $\unit{r}$ and
$\unit{\phi}$ point in the directions of increasing $r$ and $\phi$,
respectively. The self-force does not contain a component normal to
the orbital plane, and indeed, our version of the method of osculating
elements is restricted to perturbing forces that are tangent to the
plane.      

From the expression given in Eq.~(\ref{1.2}), we find that the radial 
and tangential components of the self-force are 
\begin{equation} 
R = \frac{q^2 M}{\mu} \biggl( \lambda_{\rm c} + \frac{4}{3}
\lambda_{\rm rr} \dot{r} \biggr) \frac{1}{r^3} 
\label{3.3}
\end{equation}
and 
\begin{equation} 
S = \frac{q^2 M}{\mu} \biggl( - \frac{2}{3}
\lambda_{\rm rr} \dot{\phi} \biggr) \frac{1}{r^2},  
\label{3.4}
\end{equation}
respectively. Here an overdot indicates differentiation with respect
to $t$, and time derivatives can be converted into $\phi$-derivatives
by involving Eq.~(\ref{3.2}). After differentiating Eq.~(\ref{3.1}) to
obtain $r'$, we find that Eqs.~(\ref{3.3}) and (\ref{3.4}) become 
\begin{equation} 
R = \frac{q^2 M}{\mu} \frac{(1 + ec)^3}{p^3} 
\biggl( \lambda_{\rm c} + \frac{4}{3} \lambda_{\rm rr}
\sqrt{\frac{M}{p}} es \biggr) 
\label{3.5}
\end{equation} 
and 
\begin{equation} 
S = \frac{q^2 M}{\mu} \frac{(1 + ec)^4}{p^3}  
\Biggl( -\frac{2}{3} \lambda_{\rm rr}
\sqrt{\frac{M}{p}} \Biggr),   
\label{3.6}
\end{equation} 
where $c := \cos(\phi-\omega)$ and $s := \sin(\phi-\omega)$. These
expressions are ready to be inserted within
Eqs.~(\ref{B.16})--(\ref{B.18}).  

The evolution equations come with the initial conditions    
\begin{eqnarray} 
p(\phi=0) &=:& p^*,
\nonumber \\  
e(\phi=0) &=:& e^*, 
\nonumber \\  
\omega(\phi=0) &=:& \omega^* \equiv 0, 
\nonumber \\  
t(\phi=0) &=:& t^* \equiv 0;  
\label{3.7}
\end{eqnarray} 
the values selected for $\omega^*$ and $t^*$ produce no loss of 
generality. To facilitate the integrations we introduce the
dimensionless semilatus rectum $\p$ and dimensionless time $\t$, as
well as the dimensionless parameters $\epsilon_{\rm c}$ and
$\epsilon_{\rm rr}$ that characterize the strength of the perturbing
force. These are defined by  
\begin{eqnarray} 
\p &:=& \frac{p}{p^*}, 
\nonumber \\ 
\t &:=& \sqrt{\frac{M}{p^{*3}}} t, 
\nonumber \\  
\epsilon_{\rm c} &:=& \lambda_{\rm c} \frac{q^2}{\mu p^*},
\nonumber \\  
\epsilon_{\rm rr} &:=& \frac{2}{3} \lambda_{\rm rr} 
\frac{q^2}{\mu p^*} \sqrt{\frac{M}{p^*}}. 
\label{3.8}
\end{eqnarray} 
The final form of the evolution equations is  
\begin{eqnarray} 
\p' &=& -2\epsilon_{\rm rr} \frac{1+ec}{\p^{1/2}}, 
\label{3.9} \\ 
e' &=& \epsilon_{\rm c} \frac{s(1 + ec)}{\p}
+ \epsilon_{\rm rr} \frac{(1+ec)(e - 2c - 3ec^2)}{\p^{3/2}}, \qquad 
\label{3.10} \\ 
\omega' &=& -\epsilon_{\rm c} \frac{c(1 + ec)}{e\p} 
- \epsilon_{\rm rr} \frac{(1+ec)s(2 + 3ec)}{e \p^{3/2}}, 
\label{3.11} \\ 
\t' &=& \frac{\p^{3/2}}{(1 + ec)^2}, 
\label{3.12}
\end{eqnarray} 
where $c = \cos(\phi-\omega)$ and 
$s = \sin(\phi-\omega)$. Integration proceeds from the initial values
$\p(\phi=0) =  1$, $e(\phi=0) = e^*$, $\omega(\phi=0) = 0$, and
$\t(\phi = 0) = 0$. We shall assume that $\epsilon_{\rm c}$ and
$\epsilon_{\rm rr}$ are small throughout the evolution. In spite of
the fact that $\epsilon_{\rm rr}$ is smaller than $\epsilon_{\rm c}$
by a factor of order $\sqrt{M/p^*} \ll 1$, we shall formally treat
them as being of the same order of magnitude.  

\begin{figure}
\includegraphics[angle=-90,scale=0.33]{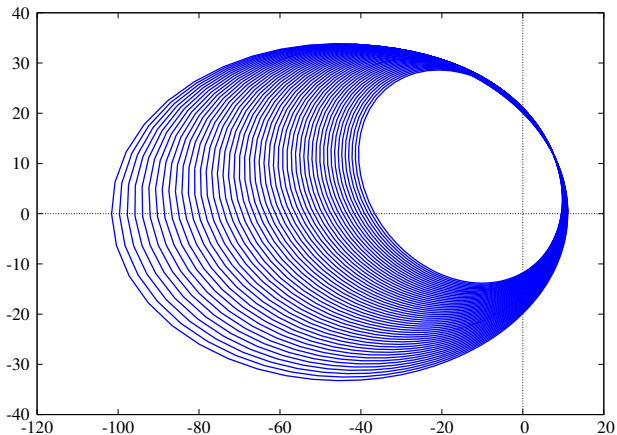}
\caption{Orbital evolution under the action of the electromagnetic
  self-force. We set $M = 1$, $q^2/\mu = 0.05$, and the orbital
  elements are integrated from the initial conditions $p^* = 20$, 
  $e^* = 0.8$, and $\omega^* = 0$. With these choices we have
  $\epsilon_{\rm c} = 2.500 \times 10^{-3}$ and 
  $\epsilon_{\rm rr} = 3.727 \times 10^{-4}$. The motion proceeds in
  the counterclockwise direction and is followed for 50 orbital cycles 
  ($0 \leq \phi < 100\pi$). A the end of the integration 
  $\epsilon_c \phi = 0.7854$. The orbit is displayed in the $x$-$y$
  plane, with $x = r\cos\phi$ and $y = r\sin\phi$. In the course of
  its evolution the orbit becomes smaller, more circular, and its
  periapsis regresses.}  
\end{figure}

Equations (\ref{3.9})--(\ref{3.12}) can easily be integrated
numerically, and the orbital motion reconstructed by inserting the
solutions within Eq.~(\ref{3.1}). The result of such a numerical 
integration is presented in Fig.~1. Our goal, however, is to obtain as
much analytical information as possible, and for this purpose we shall
construct {\it approximate solutions} to these equations, taking
advantage of the fact that $\epsilon_{\rm c}$ and 
$\epsilon_{\rm rr}$ are small. This will be carried out in the
following subsections. The approximation we shall construct is
distinct from the secular and radiative approximations considered in 
Sec.~I; we shall refer to it as the {\it multi-scale
approximation}. We shall demonstrate that our multi-scale
approximation faithfully reproduces the numerical results at all
points on the orbit, up to a time at which terms of second order in
$\epsilon_{\rm c}$ and $\epsilon_{\rm rr}$ become important. From the
multi-scale approximation we shall be able to construct secular and
radiative approximations, and we shall be able to ascertain their
accuracy.  

\subsection{Multi-scale approximation: orbital elements} 

The action of the electromagnetic self-force causes the orbital
elements $\p$, $e$, and $\omega$ to acquire a $\phi$-dependence
governed by Eqs.~(\ref{3.9})--(\ref{3.11}). The corrections to the
unperturbed solutions $\p = 1$, $e = e^*$, and $\omega = 0$ can be 
separated into two classes: {\it secular terms} that grow
monotonically with $\phi$ and {\it nonsecular terms} that oscillate
and average to zero over a complete orbital cycle ($0 \leq \phi <
2\pi$). To capture these different behaviors, we need an approximation
method that has the capability of producing a solution that stays
accurate over a long interval $0 \leq \phi < \phi_{\rm max}$, with
$\phi_{\rm max}$ of the order of $\epsilon^{-1}$, where $\epsilon$ is
the overall smallness parameter of the problem; and in this interval, 
the difference between the exact and approximate solutions must be
uniformly of order $\epsilon^2$. (Because we have introduced two such 
parameters, we shall write $\epsilon_{\rm c} = e_{\rm c} \epsilon$, 
$\epsilon_{\rm rr} = e_{\rm rr} \epsilon$ and consider $e_{\rm c}$,
$e_{\rm rr}$ to be quantities of order unity.) These requirements rule
out a simple-minded expansion in powers of $\epsilon$, because this
method would give rise to a solution that is accurate only for
$\epsilon \phi \ll 1$. We adopt instead a multi-scale analysis (see,
for example, Chapter 11 of Ref.~\cite{bender-orszag:78}).  

In a multi-scale expansion one introduces a dependence on a
``long-scale'' variable $z := \epsilon \phi$ in addition to the
dependence on the ``short-scale'' variable $\phi$. We write 
\begin{eqnarray}
\p &=& \p_0(z) + \epsilon \p_1(z,\phi) + \cdots, 
\label{3.13} \\ 
e &=& e_0(z) + \epsilon e_1(z,\phi) + \cdots, 
\label{3.14} \\ 
\omega &=& \omega_0(z) + \epsilon \omega_1(z,\phi) + \cdots,  
\label{3.15}
\end{eqnarray} 
and we seek to isolate all secular changes within the zeroth-order
quantities, and to make all first-order quantities purely
oscillatory. We use the chain rule
\[
f' = \frac{\partial f}{\partial \phi} 
+ \epsilon \frac{\partial f}{\partial z}  
\]
to evaluate the total derivative with respect to $\phi$ of a function 
$f(z,\phi)$. 

\begin{widetext} 
To proceed we substitute Eqs.~(\ref{3.13})--(\ref{3.15}) into 
Eqs.~(\ref{3.9})--(\ref{3.11}) and obtain, to first order in
$\epsilon$, 
\begin{eqnarray} 
\frac{d \p_0}{dz} + \frac{\partial \p_1}{\partial \phi} &=& 
- \frac{2 e_{\rm rr}}{\p_0^{1/2}} \Bigl[ 1 + e_0 \cos v 
    \bigr], 
\label{3.16} \\ 
\frac{d e_0}{dz} + \frac{\partial e_1}{\partial \phi} &=& 
\frac{e_{\rm c}}{\p_0} \biggl[ \sin v 
+ \frac{1}{2} e_0 \sin 2v \biggr] 
\nonumber \\ & & \mbox{} 
- \frac{e_{\rm rr}}{\p_0^{3/2}} \biggl[ \frac{3}{2} e_0 
+ \frac{1}{4} (8 + 5 e_0^2) \cos v 
+ \frac{5}{2} e_0 \cos 2v 
+ \frac{3}{4} e_0^2 \cos 3v \biggr], 
\label{3.17} \\ 
\frac{d \omega_0}{dz} + \frac{\partial \omega_1}{\partial \phi} &=&  
-\frac{e_{\rm c}}{2 \p_0} \biggl[ 1 
+ \frac{2}{e_0} \cos v + \cos 2v \biggr] 
\nonumber \\ & & \mbox{} 
- \frac{e_{\rm rr}}{\p_0^{3/2}} \biggr[ 
\frac{8 + 3e_0^2}{4 e_0} \sin v 
+ \frac{5}{2} \sin 2v 
+ \frac{3}{4} e_0 \sin 3v \biggr],  
\label{3.18}
\end{eqnarray} 
where $v := \phi - \omega_0$. It is easy to recognize the terms on the
right-hand sides that drive the secular changes in the orbital
elements. We isolate these changes by setting 
\begin{equation} 
\frac{d \p_0}{dz} = - \frac{2 e_{\rm rr}}{\p_0^{1/2}}, 
\qquad \qquad
\frac{d e_0}{dz} = - \frac{3 e_{\rm rr} e_0}{2 \p_0^{3/2}}, 
\qquad \qquad 
\frac{d \omega_0}{dz} = -\frac{e_{\rm c}}{2 \p_0}. 
\label{3.19} 
\end{equation} 
The nonsecular (oscillatory) corrections are then obtained by
integrating   
\begin{eqnarray} 
\frac{\partial \p_1}{\partial \phi} &=& 
- \frac{2 e_{\rm rr} e_0}{\p_0^{1/2}} \cos v,  
\label{3.20} \\ 
\frac{\partial e_1}{\partial \phi} &=& 
\frac{e_{\rm c}}{\p_0} \biggl[ \sin v 
+ \frac{1}{2} e_0 \sin 2v \biggr] 
\nonumber \\ & & \mbox{} 
- \frac{e_{\rm rr}}{\p_0^{3/2}} \biggl[ 
\frac{1}{4} (8 + 5 e_0^2) \cos v 
+ \frac{5}{2} e_0 \cos 2v 
+ \frac{3}{4} e_0^2 \cos 3v \biggr], 
\label{3.21} \\ 
\frac{\partial \omega_1}{\partial \phi} &=& 
-\frac{e_{\rm c}}{2 \p_0} \biggl[ 
\frac{2}{e_0} \cos v + \cos 2v \biggr] 
\nonumber \\ & & \mbox{} 
- \frac{e_{\rm rr}}{\p_0^{3/2}} \biggr[ 
\frac{8 + 3e_0^2}{4 e_0} \sin v 
+ \frac{5}{2} \sin 2v 
+ \frac{3}{4} e_0 \sin 3v \biggr]. 
\label{3.22}
\end{eqnarray} 
We must impose the initial conditions $\p_0 + \epsilon \p_1 + \cdots  
= 1$, $e_0 + \epsilon e_1 + \cdots = e^*$, and $\omega_0 + \epsilon
\omega_1 + \cdots = 0$ when $\phi = 0$. 
\end{widetext} 
 
The general solutions to Eqs.~(\ref{3.19}) are  
\begin{eqnarray} 
\p_0 &=& a \bigl( 1 - 3 e_{\rm rr} z / a^{3/2} \bigr)^{2/3}, 
\label{3.23} \\ 
e_0 &=& b \bigl( 1 - 3 e_{\rm rr} z / a^{3/2} \bigr)^{1/2}, 
\label{3.24} \\ 
\omega_0 &=& c - \frac{e_{\rm c} a^{1/2}}{2 e_{\rm rr}} 
\Bigl[ 1 - \bigl( 1 - 3 e_{\rm rr} z / a^{3/2} \bigr)^{1/3} \Bigr], 
\label{3.25}
\end{eqnarray}
where $a$, $b$, and $c$ are constants of integration that will be
determined. 

The purely oscillatory solutions to Eqs.~(\ref{3.20})--(\ref{3.22})
are 
\begin{eqnarray} 
\epsilon \p_1 &=& 
- \frac{2 \epsilon_{\rm rr} e_0}{\p_0^{1/2}} \sin v,  
\label{3.26} \\ 
\epsilon e_1 &=& 
- \frac{\epsilon_{\rm c}}{\p_0} \biggl[ \cos v 
+ \frac{1}{4} e_0 \cos 2v \biggr] 
\nonumber \\ & & \mbox{} 
- \frac{\epsilon_{\rm rr}}{4 \p_0^{3/2}} \biggl[ 
(8 + 5 e_0^2) \sin v 
+ 5 e_0 \sin 2v
\nonumber \\ & & \hspace*{40pt} \mbox{} 
+ e_0^2 \sin 3v \biggr], 
\label{3.27} \\ 
\epsilon \omega_1 &=& 
-\frac{\epsilon_{\rm c}}{\p_0} \biggl[ 
\frac{1}{e_0} \sin v 
+ \frac{1}{4} \sin 2v \biggr] 
\nonumber \\ & & \mbox{} 
+ \frac{\epsilon_{\rm rr}}{4 \p_0^{3/2}} \biggr[ 
\frac{8 + 3e_0^2}{e_0} \cos v 
+ 5 \cos 2v 
\nonumber \\ & & \hspace*{40pt} \mbox{} 
+ e_0 \cos 3v \biggr]. 
\label{3.28}
\end{eqnarray} 
We recall that $v = \phi - \omega_0$. 

To relate the constants $a$, $b$, and $c$ to the initial conditions we
note that $\p_0(0) = a$ and $\epsilon \p_1(0) = 0$; we therefore have
$a = \p^* = 1$. Similarly, we note that $e_0(0) = b$ and $\epsilon
e_1(0) = -\frac{1}{4} \epsilon_{\rm c} (4 + b)$; we therefore have 
$b = e^* + \frac{1}{4} \epsilon_{\rm c} (4 + e^*) 
+ O(\epsilon^2)$. Finally, we note that $\omega_0(0) = c$ and
$\epsilon \omega_1(0) = \epsilon_{\rm rr} (8 + 5 b + 4 b^2)/(4 b)$; we 
therefore have $c = -\epsilon_{\rm rr} (8 + 5 e^* + 4 e^{*2})/(4 e^*) 
+ O(\epsilon^2)$. Making these substitutions in 
Eqs.~(\ref{3.22})--(\ref{3.24}) gives  
\begin{eqnarray} 
\p_0 &=& \bigl( 1 - 3 \epsilon_{\rm rr} \phi \bigr)^{2/3}, 
\label{3.29} \\ 
e_0 &=& e^* \biggl[ 1 + \epsilon_{\rm c} \frac{4 + e^*}{4 e^*}
  \biggr] \bigl( 1 - 3 \epsilon_{\rm rr} \phi \bigr)^{1/2}, 
\label{3.30} \\ 
\omega_0 &=& -\epsilon_{\rm rr} \frac{8 + 5 e^* + 4 e^{*2}}{4 e^*} 
\nonumber \\ & & \mbox{} 
- \frac{\epsilon_{\rm c}}{2 \epsilon_{\rm rr}} 
\Bigl[ 1 - \bigl( 1 - 3 \epsilon_{\rm rr} \phi \bigr)^{1/3} \Bigr].  
\label{3.31}
\end{eqnarray}
These expressions describe the secular changes in the orbital 
elements. Equations (\ref{3.26})--(\ref{3.28}), on the other hand,
describe the nonsecular (oscillatory) changes. All together, these
results give us the desired multi-scale approximation for the orbital
elements.\footnote{We take this opportunity to make an observation. We
  notice from Eq.~(\ref{3.29}) that when $\epsilon \phi$ is comparable 
  to unity, the dissipative term in the self-force produces a change
  in $\p$ that is also of order unity. In this calculation the
  radiation-reaction force is linear in $\epsilon$, and there are no
  corrections of order $\epsilon^2$. If such corrections were present,
  however, they would produce an additional change of order $\epsilon$
  in $\p$. Next we notice from Eq.~(\ref{3.30}) that the conservative
  term in the self-force produces a change of order $\epsilon$ in $e$,
  in addition to the change of order unity that comes from the
  radiation-reaction force. We conclude that {\it second-order terms
  in the radiation-reaction force} would produce effects that scale
  with the same power of $\epsilon$ as those produced by the
  conservative force. As we shall see below, the conservative force
  must be included in the calculation when the evolution of the
  orbital phase is required to stay accurate to order $\epsilon^0$
  during a radiation-reaction time. In a context where the
  radiation-reaction force would contain a second-order term, the 
  same accuracy would be achieved only after including this term as
  well in the calculation. We thank Tanja Hinderer and \'Eanna
  Flanagan for making this point clear to us.}    
We compare the approximations with exact numerical results in Figs.~2,
3, and 4.    

\begin{figure}
\includegraphics[angle=-90,scale=0.33]{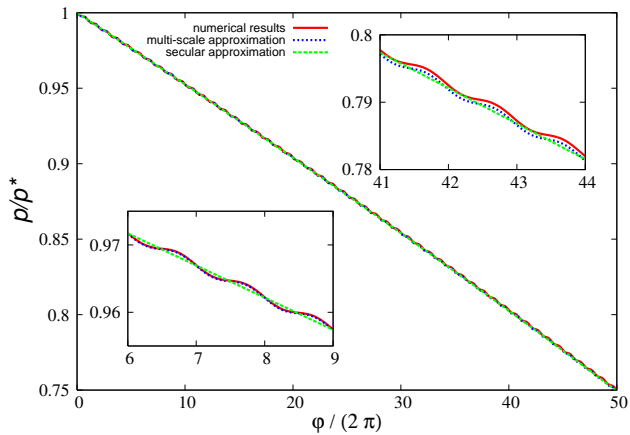}
\caption{Multi-scale approximation for $p(\phi/2\pi)$ compared with
  exact numerical results. The numerical conditions are the same as in
  Fig.~1. The solid curve in red shows the exact evolution as computed
  numerically. The dotted curve in blue shows the evolution as
  predicted by the multi-scale approximation, which includes a secular
  term as well as oscillations. The dashed curve in green shows the
  secular piece of the multi-scale approximation. The large panel shows
  the entire evolution from $\phi = 0$ to $\phi = 100\pi$. The first
  inset (bottom left) shows the evolution in the small interval $6 <
  \phi/(2\pi) < 9$; early in the evolution the multi-scale
  approximation is extremely accurate. The second inset (top right)
  shows the evolution in the small interval $41 < \phi/(2\pi) < 44$;
  here the multi-scale approximation is less accurate, because
  $\epsilon \phi$ has become comparable to unity.} 
\end{figure}

\begin{figure}
\includegraphics[angle=-90,scale=0.33]{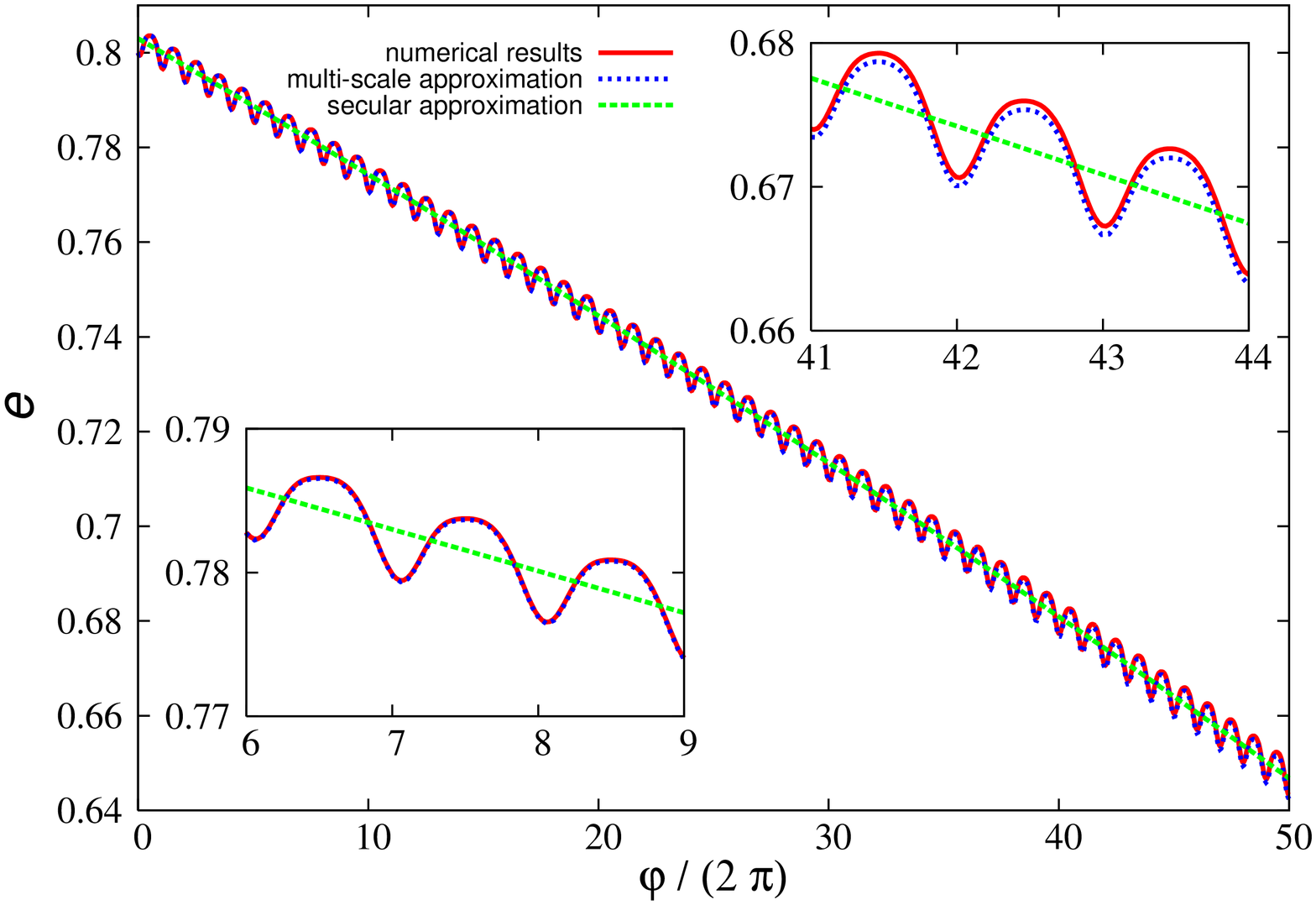}
\caption{Multi-scale approximation for $e(\phi/2\pi)$ compared with
  exact numerical results. The numerical conditions are the same as in
  Fig.~1, and the caption of Fig.~2 provides the relevant details.}  
\end{figure}

\begin{figure}
\includegraphics[angle=-90,scale=0.33]{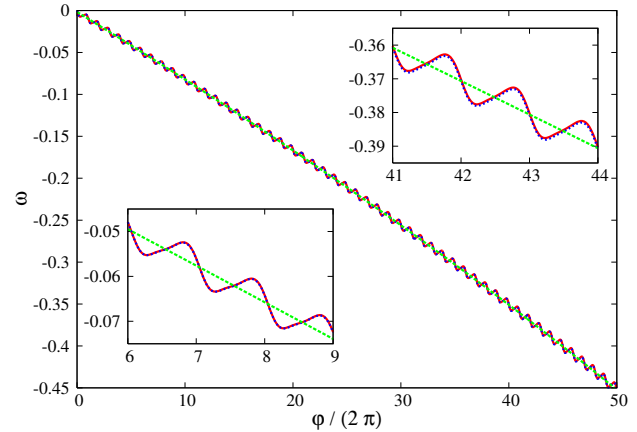}
\caption{Multi-scale approximation for $\omega(\phi/2\pi)$ compared
  with exact numerical results. The numerical conditions are the same
  as in Fig.~1, and the caption of Fig.~2 provides the relevant 
  details.}   
\end{figure}

\subsection{Multi-scale approximation: time}

We now wish to construct a multi-scale approximation for the time
function $\t(\phi)$. To begin we recall that Eq.~(\ref{3.12}) must be  
integrated numerically even in the unperturbed situation, when $p$,
$e$, and $\omega$ are all constant; the approximation, therefore, will  
also involve a numerical integration. A lazy option presents itself:
The time function could be obtained simply by inserting our
multi-scale approximations for $\p(\phi)$, $e(\phi)$, and
$\omega(\phi)$ within Eq.~(\ref{3.12}) and performing the integration 
numerically. In an effort to obtain maximum analytical insight,
however, we choose to proceed differently.  

We substitute Eqs.~(\ref{3.13})--(\ref{3.15}) and the explicit
expressions of Eqs.~(\ref{3.26})--(\ref{3.28}) into
Eq.~(\ref{3.12}), and we expand in powers of $\epsilon$. Through first
order we obtain  
\begin{eqnarray} 
\t' &=& \frac{\p_0^{3/2}}{(1 + e_0 \cos v)^2} 
\nonumber \\ & & \mbox{} 
+ \frac{1}{2} \epsilon_{\rm c} \p_0^{1/2} 
\frac{4 + e_0\cos v}
     {(1 + e_0\cos v)^3} 
\nonumber \\ & & \mbox{} 
- \frac{1}{2} \epsilon_{\rm rr} e_0 
\frac{\sin v + e_0 \sin 2v} 
     {(1 + e_0\cos v)^3}, 
\label{3.32}
\end{eqnarray}
where $v = \phi-\omega_0$, and $\p_0$, $e_0$, and $\omega_0$ are the
functions of $z := \epsilon_{\rm rr} \phi$ displayed in 
Eqs.~(\ref{3.29})--(\ref{3.31}). The first line in Eq.~(\ref{3.32}) is
obtained by inserting $\p = \p_0$, $e=e_0$, and $\omega = \omega_0$
within Eq.~(\ref{3.12}); the second and third lines are contributed by
the oscillatory terms $\p_1$, $e_1$, and $\omega_1$. 

We wish to find an approximate solution to Eq.~(\ref{3.32}), and once
more we wish to distinguish between secular and nonsecular terms. Two
sources of complications present themselves. First, while it was easy
in Eq.~(\ref{3.16})--(\ref{3.18}) to separate the secular terms from
the oscillations, the factors of $1 + e_0\cos v$ in the denominators
of Eq.~(\ref{3.32}) make this separation more difficult. Second,
while $\p$, $e$, and $\omega$, are simply constant at the unperturbed 
level, the Keplerian version of $\t(\phi)$ is already a complicated  
function of $\phi$ that contains secular and oscillating terms. The
situation here is therefore more complicated, but we will,
nevertheless, be able to express the solution to Eq.~(\ref{3.32}) in
the form   
\begin{equation} 
\t(\phi) = \t_0(\phi) + \epsilon \t_1(\phi) + \cdots, 
\label{3.33}
\end{equation} 
with $\t_0(\phi)$ incorporating the Keplerian behavior (including 
secular terms and oscillations) as well as the secular changes
produced by the electromagnetic self-force, and with 
$\epsilon \t_1(\phi)$ being purely oscillatory.   

To isolate the oscillatory terms in Eq.~(\ref{3.32}) we calculate the
averages 
\[
\langle f \rangle_\phi(\phi) := 
\frac{1}{2\pi} \int_{\phi-\pi}^{\phi+\pi} f(\phi')\, d\phi'  
\]
of the various functions of $\phi$ that appear on its right-hand
side; these averages are calculated while keeping $\p_0$, $e_0$, and 
$\omega_0$ constant over the integration domain. Defining 
$f_1 = (1+e_0\cos v)^{-2}$, 
$f_2 = (4 + e_0\cos v) (1+e_0\cos v)^{-3}$, and 
$f_3 = (\sin v + e_0 \sin 2v)(1+e_0\cos v)^{-3}$, we find that  
$\langle f_1 \rangle_\phi = (1-e_0^2)^{-3/2}$, 
$\langle f_2 \rangle_\phi = \frac{1}{2}(8+e_0^2)(1-e_0^2)^{-5/2}$, and  
$\langle f_3 \rangle = 0$. The function multiplying 
$\epsilon_{\rm rr}$ in Eq.~(\ref{3.32}) is therefore purely
oscillatory, but the function multiplying $\epsilon_{\rm c}$ contains
a secular component. To isolate this we rewrite Eq.~(\ref{3.32}) into
the equivalent form  
\begin{eqnarray} 
\t' &=& \frac{\p_0^{3/2}}{(1 + e_0 \cos v)^2} 
\biggl[1 + \epsilon_{\rm c} \frac{8+e_0^2}{4\p_0 (1-e_0^2)} \biggr] 
\nonumber \\ & & \mbox{} 
- \frac{1}{4} \epsilon_{\rm c} \frac{\p_0^{1/2} e_0}{1-e_0^2}  
\frac{9 e_0 + 3(2+e_0^2) \cos v} 
     {(1 + e_0\cos v)^3}  
\nonumber \\ & & \mbox{} 
- \frac{1}{2} \epsilon_{\rm rr} e_0 
\frac{\sin v + e_0 \sin 2v}{\bigl(1 + e_0\cos v)^3}, 
\label{3.34}
\end{eqnarray}
in which a term $\frac{1}{4} \epsilon_{\rm c} \p_0^{1/2} (8+e_0^2)
(1-e_0^2)^{-1} (1 + e_0 \cos v)^{-2}$ was removed from the second line
in Eq.~(\ref{3.32}) and inserted within the first line. In
Eq.~(\ref{3.34}), the functions that appear in the second and third
lines are purely oscillatory.  

It is important to notice that the term proportional to 
$\epsilon_{\rm c}$ in the first line of Eq.~(\ref{3.34}) is a secular
correction to $\t'$ that originates with the oscillatory terms $\p_1$,
$e_1$, and $\omega_1$ in the orbital elements. These oscillations
combine in a nonlinear fashion, and they contribute an additional
secular term beyond the one that comes from $\p_0$, $e_0$, and
$\omega_0$. It would be a significant mistake to discard the
oscillations in the orbital elements when constructing the time
function.  

The solution to Eq.~(\ref{3.34}) is 
\begin{eqnarray} 
\t &=& \int_0^\phi 
\frac{\p_0^{3/2}}{(1 + e_0 \cos v')^2}  
\biggl[1 + \epsilon_{\rm c} \frac{8+e_0^2}{4\p_0 (1-e_0^2)} \biggr]\,
d\phi' 
\nonumber \\ & & \mbox{} 
- \frac{1}{4} \epsilon_{\rm c} \int_0^\phi 
\frac{\p_0^{1/2} e_0}{1-e_0^2} \frac{9 e_0 + 3(2+e_0^2) \cos v'} 
     {(1 + e_0\cos v')^3}\, d\phi'   
\nonumber \\ & & \mbox{} 
- \frac{1}{2} \epsilon_{\rm rr} \int_0^\phi 
e_0 \frac{\sin v' + e_0 \sin 2v'}{\bigl(1 + e_0\cos v')^3}\, 
d\phi', 
\label{3.35}
\end{eqnarray}
where $v' := \phi' - \omega_0(\phi')$. We must leave the first
integral alone, but we shall manage to evaluate the second and third 
integrals. Because the changes in $\p_0$, $e_0$, and $\omega_0$ are of 
order $\epsilon$, because the second and third integrals already come
with a factor of $\epsilon$ in front, because the integrands are
purely oscillatory functions, and because the calculation of
$\t(\phi)$ is carried out consistently to first order in $\epsilon$,
we are permitted to treat $\p_0$, $e_0$, and $\omega_0$ as constants
when evaluating the integrals. We thus obtain 
\begin{eqnarray} 
\t &=& \int_0^\phi 
\frac{\p_0^{3/2}}{(1 + e_0 \cos v')^2}  
\biggl[1 + \epsilon_{\rm c} \frac{8+e_0^2}{4\p_0 (1-e_0^2)} \biggr]\,
d\phi' 
\nonumber \\ & & \mbox{} 
- \frac{1}{2} \epsilon_{\rm c} \frac{\p_0^{1/2} e_0}{1-e_0^2} 
\frac{3\sin v + \frac{3}{4} e_0 \sin 2v}{(1 + e_0\cos v)^2}   
\nonumber \\ & & \mbox{} 
- \frac{1}{4} \epsilon_{\rm rr} \biggl[ 
\frac{3 + 4 e_0 \cos v}{(1 + e_0\cos v)^2} 
- \frac{3 + 4 e_0}{(1+e_0)^2} \biggr].
\label{3.36}
\end{eqnarray}
We have not yet achieved the form of Eq.~(\ref{3.33}). The reason is
that while the function in the second line of Eq.~(\ref{3.36}) is
purely oscillatory (it has a zero average), this is not true of the
function in the third line. Defining $f_4 = (3+4 e_0\cos v)
(1 + e\cos v)^{-2}$, we find that $\langle f_4 \rangle_\phi  
= (3-4e_0^2)(1-e_0^2)^{-3/2}$ and this combines with the second term 
on the third line to contribute secular terms. Removing these from the
third line and inserting them within the first line, we finally arrive
at the desired expression for the time function $\t(\phi)$.  

Our final result is that $\t(\phi)$ can be expressed as in
Eq.~(\ref{3.33}), with 
\begin{eqnarray}  
\t_0 &=& \int_0^\phi 
\frac{\p_0^{3/2}}{(1 + e_0 \cos v')^2}  
\biggl[1 + \epsilon_{\rm c} \frac{8+e_0^2}{4\p_0 (1-e_0^2)} \biggr]\, 
d\phi' 
\nonumber \\ & & \mbox{} 
+ \frac{1}{4} \epsilon_{\rm rr} \biggl[ \frac{3+4e_0}{(1+e_0)^2} 
- \frac{3-4 e_0^2}{(1-e_0^2)^{3/2}} \biggr] 
\label{3.37}
\end{eqnarray} 
and 
\begin{eqnarray} 
\epsilon \t_1 &=& 
-\frac{1}{2} \epsilon_{\rm c} \frac{\p_0^{1/2} e_0}{1-e_0^2}  
\frac{3\sin v + \frac{3}{4} e_0 \sin 2v}{(1 + e_0\cos v')^2}   
\nonumber \\ & & \mbox{} 
- \frac{1}{4} \epsilon_{\rm rr} \biggl[ 
\frac{3 + 4 e_0 \cos v}{(1 + e_0\cos v)^2} 
- \frac{3 - 4 e_0^2}{(1-e_0^2)^{3/2}} \biggr], \qquad 
\label{3.38}
\end{eqnarray} 
where $v = \phi - \omega_0$, and $\p_0$, $e_0$, and $\omega_0$ are the 
functions of $z := \epsilon_{\rm rr} \phi$ displayed in 
Eqs.~(\ref{3.29})--(\ref{3.31}). By design, $\t_0(\phi)$ incorporates
the Keplerian behavior (including all Keplerian oscillations) in
addition to the secular changes produced by the perturbing force; the
function $\epsilon \t_1(\phi)$ is purely oscillatory, in the sense
that its $\phi$-average is zero. A comparison between the exact time
function $t(\phi)$ and the multi-scale approximation is presented in
Fig.~5.   

\begin{figure}
\includegraphics[angle=-90,scale=0.33]{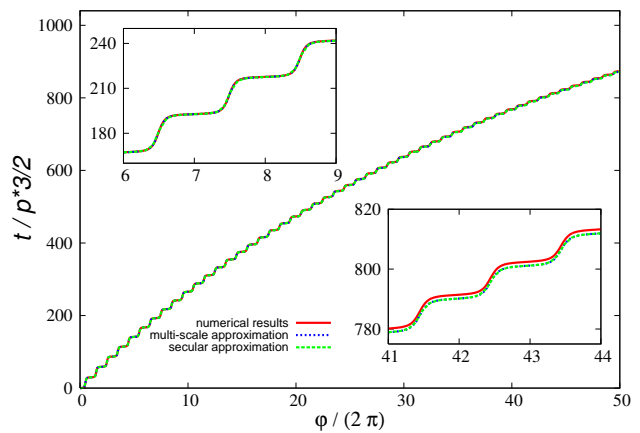}
\caption{Multi-scale approximation for $t(\phi/2\pi)$ compared with
  exact numerical results. The numerical conditions are the same as in 
  Fig.~1. The solid curve in red shows the exact evolution as computed
  numerically. The dotted curve in blue shows the evolution as
  predicted by the multi-scale approximation $t_0 + \epsilon t_1$,
  where $t_0$ incorporates the Keplerian behavior in addition to the
  secular changes produced by the perturbing force, and where 
  $\epsilon t_1(\phi)$ is purely oscillatory. The dashed curve in
  green is a plot of $t_0$ only. The large panel shows the entire
  evolution from $\phi = 0$ to $\phi = 100\pi$. The first inset (top
  left) shows the evolution in the small interval $6 < \phi/(2\pi) <
  9$. The second inset (bottom right) shows the evolution in the 
  interval $41 < \phi/(2\pi) < 44$.}
\end{figure}

\section{Discussion} 

With the technical details out of the way, we may now return to the
themes introduced in Sec.~I. To launch the discussion we summarize the
main results obtained in Sec.~III. 

\subsection{Summary of our results} 

In the method of osculating orbital elements, the motion of a charged
particle subjected to the electromagnetic self-force of
Eq.~(\ref{1.2}) is at all times described by 
\begin{equation} 
r(\phi) = \frac{p}{1 + e\cos(\phi - \omega)}, 
\label{4.1}
\end{equation} 
the Keplerian relation of Eq.~(\ref{3.1}). The orbital elements $p$,
$e$, $\omega$, however, acquire a $\phi$-dependence that accounts for
the perturbation created by the self-force. With the definitions of
Eq.~(\ref{3.8}), these quantities evolve according to
Eqs.~(\ref{3.9})--(\ref{3.11}), and integrating Eq.~(\ref{3.12})
produces $t(\phi)$. The motion is then fully determined.  

The evolution equations can be integrated numerically, or they can be
integrated analytically via a multi-scale analysis that produces a
faithful approximation over the long interval $0 \leq \phi \alt
\epsilon^{-1}$. Moreover, the multi-scale analysis produces a clean
separation of the solutions into secular and oscillatory pieces. The
secular changes in the orbital elements are described by the functions
$p_0$, $e_0$, and $\omega_0$ displayed in
Eqs.~(\ref{3.29})--(\ref{3.31}). We copy them here for convenience:  
\begin{eqnarray} 
p_0 &=& p^* \bigl( 1 - 3 \epsilon_{\rm rr} \phi \bigr)^{2/3}, 
\label{4.2} \\ 
e_0 &=& e^* \biggl[ 1 + \epsilon_{\rm c} \frac{4 + e^*}{4 e^*}
  \biggr] \bigl( 1 - 3 \epsilon_{\rm rr} \phi \bigr)^{1/2}, 
\label{4.3} \\ 
\omega_0 &=& -\epsilon_{\rm rr} \frac{8 + 5 e^* + 4 e^{*2}}{4 e^*} 
\nonumber \\ & & \mbox{} 
- \frac{\epsilon_{\rm c}}{2 \epsilon_{\rm rr}} 
\Bigl[ 1 - \bigl( 1 - 3 \epsilon_{\rm rr} \phi \bigr)^{1/3} \Bigr],   
\label{4.4}
\end{eqnarray}
where 
\begin{eqnarray} 
\epsilon_{\rm c} &:=& \lambda_{\rm c} \frac{q^2}{\mu p^*},
\label{4.5} \\
\epsilon_{\rm rr} &:=& \frac{2}{3} \lambda_{\rm rr} 
\frac{q^2}{\mu p^*} \sqrt{\frac{M}{p^*}},  
\label{4.6}
\end{eqnarray} 
and where $p^* := p(\phi=0)$ and $e^* := e(\phi=0)$; we have set
$\omega(\phi=0) = 0$. The oscillatory changes in the orbital elements
are given by $\epsilon p_1$, $\epsilon e_1$, and $\epsilon\omega_1$
displayed in Eqs.~(\ref{3.26})--(\ref{3.28}). 

The piece of the time function that incorporates Keplerian behavior  
and secular changes produced by the electromagnetic self-force is
$t_0$, and this is displayed in Eq.~(\ref{3.37}). We copy it here for
convenience:  
\begin{eqnarray}  
t_0 &=& \sqrt{\frac{p^{*3}}{M}} \Biggl\{ \int_0^\phi 
\frac{(p_0/p^*)^{3/2}}{(1 + e_0 \cos v')^2}  
\biggl[1 + \epsilon_{\rm c} \frac{(8+e_0^2) p^*}{4 p_0 (1-e_0^2)}
  \biggr]\, d\phi' 
\nonumber \\ & & \mbox{} 
+ \frac{1}{4} \epsilon_{\rm rr} \biggl[ \frac{3+4e_0}{(1+e_0)^2} 
- \frac{3-4 e_0^2}{(1-e_0^2)^{3/2}} \biggr] \Biggr\}, 
\label{4.7}
\end{eqnarray} 
where $v' = \phi' - \omega_0(\phi')$; this, like the Keplerian time
function, is expressed in terms of an integral that must be evaluated
numerically. The oscillatory piece of the time function is $\epsilon
t_1$, and this is given by Eq.~(\ref{3.38}). We recall that the term
proportional to $\epsilon_{\rm c}$ inside the integral is produced by 
the oscillatory pieces $p_1$, $e_1$, and $\omega_1$ of the orbital
elements; the oscillations combine to produce a secular correction 
to the time function.   

\subsection{Conservative and dissipative terms in the self-force} 

The effect of each term in Eq.~(\ref{1.2}) can easily be identified if
we focus our attention on the secular changes in the orbital elements
and time function described by Eqs.~(\ref{4.2})--(\ref{4.7}). Because
these accumulate in the long run, while the oscillations that are not
contained in those equations average to zero, it is clear that it is
the secular pieces of $p(\phi)$, $e(\phi)$, $\omega(\phi)$, and
$t(\phi)$ that are the most important to capture. 

The effects of the conservative piece of the self-force are identified
by selecting the terms in $\epsilon_{\rm c}$; the effects of the
radiation-reaction piece are identified by $\epsilon_{\rm rr}$. An
examination of Eqs.~(\ref{4.2})--(\ref{4.7}) reveals that the
radiation-reaction force drives secular changes in the principal
orbital elements $p$ and $e$, but that it affects $\omega$ only
indirectly, and only if there is a conservative force. In addition,
the radiation-reaction force affects the time function indirectly
through the changes in the principal elements, and also directly as
can be seen in the second line of Eq.~(\ref{4.7}). On the other hand,
the conservative force drives secular changes in the positional
element $\omega$, but it affects $e$ only through the factor that
comes in front of $(1-3\epsilon_{\rm rr} \phi)^{1/2}$. In addition,
the conservative force affects the time function directly, as can be
seen from the correction term inside the integral.  

The combined effects of the conservative and radiation-reaction forces
on the time function can be neatly summarized by computing 
$P := \int_\phi^{\phi + 2\pi} t(\phi')\, d\phi'$, the period of an
orbital cycle. Ignoring the changes in the orbital elements while
performing the integration, we obtain 
\begin{equation} 
P = 2\pi \sqrt{ \frac{p_0^3}{M(1-e_0^2)^3} } \biggl( 1 
+ \frac{1}{4} \epsilon_{\rm c} \frac{p^*}{p_0} \frac{8+e_0^2}{1-e_0^2} 
\biggr).  
\label{4.8}
\end{equation} 
The factor in front of the large brackets is the Keplerian period
expressed in terms of the changing orbital elements; these changes, we
recall, are driven by the radiation-reaction force. The second term
gives the correction contributed by the conservative force. It may be
recalled that this correction originates from oscillations in $p$,
$e$, and $\omega$, and it may be noted that it becomes large when $e_0
\to 1$.   

\subsection{Limitations of the radiative approximation} 

As we have defined it in Sec.~I, the radiative approximation is
obtained by setting $\epsilon_{\rm c} = 0$ in our results. This
preserves the secular changes in $p$ and $e$, but it completely turns
off the secular evolution of $\omega$. In addition, the radiative
approximation discards an important correction term in the time
function, the one proportional to $\epsilon_{\rm c}$ in
Eq.~(\ref{4.7}). This term, in fact, dominates over the
radiation-reaction corrections, because $\epsilon_{\rm c}$ is
numerically larger than $\epsilon_{\rm rr}$ by a factor of order
$\sqrt{p^*/M} \gg 1$, as can be seen from Eqs.~(\ref{4.5}) and
(\ref{4.6}). In addition, we have noted that the correction becomes
increasingly large as $e_0$ increases toward unity. As a result, the
radiative approximation provides a poor estimation of the orbital
period, as can be seen from Eq.~(\ref{4.8}).   

The combined effect of omitting the secular changes in $\omega$ and
missing an important correction in the time function can be seen most
clearly by examining the radial phase variable $\Phi := \phi - \omega$ 
expressed as a function of time; this is the function that appears in
$r(t) = p/(1 + e\cos\Phi)$, as can be seen from Eq.~(\ref{4.1}). 
In Fig.~6 we compare the results of two numerical computations, one
carried out with the full electromagnetic self-force (including
conservative and radiation-reaction pieces), the other carried out in
the radiative approximation, with only the radiation-reaction piece of
the self-force. The figure, and the quantitative analysis presented in
the caption, reveal very clearly that the radiative approximation
gives a rather poor representation of the phase function; in our
simulation, the mismatch after 50 orbits is nearly three full radial
cycles. The origin of the discrepancy is also clearly identified in 
the caption: It is the missing conservative correction to the time
function $t(\phi)$ that is mostly responsible for the phase mismatch.  

\begin{figure}
\includegraphics[angle=-90,scale=0.33]{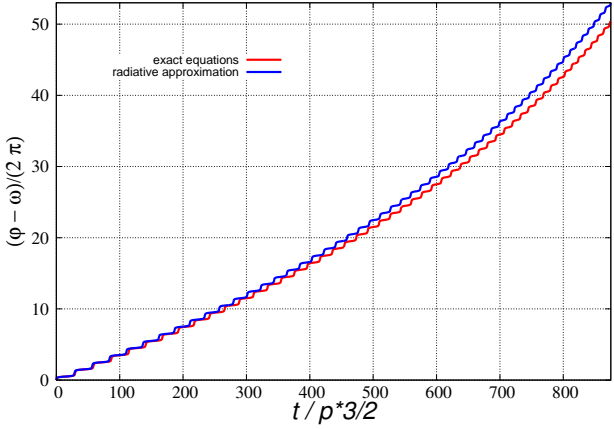}
\caption{Effect of the radiative approximation on the radial phase
  function $\Phi(t)$, where $\Phi = \phi - \omega(\phi)$. Plotted are 
  $\Phi/(2\pi)$ versus $\t = \sqrt{M/p^{*3}} t$, with the same
  numerical conditions as in Fig.~1. The lower curve in red is the
  exact evolution as computed numerically with the help of the full 
  self-force, which includes conservative and radiation-reaction
  pieces. The upper curve in blue is the evolution obtained in the
  radiative approximation, in which the conservative force is switched
  off. At the end of the integration, for $\t = 875$, we have
  $\Phi/(2\pi) = 50.40$ under the action of the full self-force, with
  a contribution $\omega/(2\pi) = -0.073$ coming from the periapsis 
  advance. The radiative approximation gives instead $\Phi/(2\pi) =
  53.06$, with a contribution $\omega/(2\pi) = 0.00016$ coming from
  the shift in periapsis. The total phase mismatch is $\Delta \Phi =
  2.66(2\pi)$, nearly three radial cycles out of fifty orbits. Because
  the contribution from $\omega$ is small in both cases, we conclude
  that the main source of error is contained in the missing
  conservative terms in $t(\phi)$.} 
\end{figure}

\subsection{Secular approximation: $\phi$-average} 

The multi-scale approximation method of Sec.~III was adopted precisely
because it produces a clean separation between secular and nonsecular
terms in the expressions for the orbital elements and the time
function. The zeroth-order quantities displayed earlier in this
section were constructed so as to represent the secular changes, and
the oscillatory corrections were carefully designed to average to
zero. It is clear, therefore, that Eqs.~(\ref{4.2})--(\ref{4.7})
achieve the goals of a secular approximation, and we would be
justified to write 
\begin{eqnarray} 
p_{\rm sec}(\phi) = p_0(\phi), \qquad
e_{\rm sec}(\phi) = e_0(\phi),  
\nonumber \\ 
\omega_{\rm sec}(\phi) = \omega_0(\phi), \qquad
t_{\rm sec}(\phi) = t_0(\phi). 
\label{4.9}
\end{eqnarray}
This, in the language introduced in Sec.~II, consists of defining the
secular approximation by averaging the exact solution (as
represented by the multi-scale approximation, which has been
demonstrated to be faithful to the exact numerical results) over the
orbital parameter $\phi$. 

The question we wish to explore here is whether the secular
approximation of Eq.~(\ref{4.9}) could be formulated directly, without
the help of the exact solution. The dynamical equations that govern 
the secular approximation can be obtained by differentiating 
Eqs.~(\ref{4.2})--(\ref{4.7}) with respect to $\phi$. We obtain 
\begin{eqnarray} 
p'_{\rm sec} &=& -2 (\epsilon_{\rm rr} p^{*3/2})\, p_{\rm sec}^{-1/2},  
\label{4.10} \\ 
e'_{\rm sec} &=& -\frac{3}{2} (\epsilon_{\rm rr} p^{*3/2})\,  
  e_{\rm sec}\, p_{\rm sec}^{-3/2}, 
\label{4.11} \\ 
\omega'_{\rm sec} &=& -\frac{1}{2} (\epsilon_{\rm c} p^*)\,  
  p_{\rm sec}^{-1}, 
\label{4.12} \\ 
t'_{\rm sec} &=& \frac{\sqrt{p_{\rm sec}^3/M}}
  {(1 + e_{\rm sec} \cos v)^2}    
\biggl[1 + \frac{(\epsilon_{\rm c} p^*)(8+e_{\rm sec}^2)}
  {4 (1-e_{\rm sec}^2) p_{\rm sec}} \biggr], \qquad 
\label{4.13}
\end{eqnarray} 
in which $v = \phi - \omega_{\rm sec}(\phi)$ and where we have ignored 
(as we should) terms of order $\epsilon^2$. These equations must come
with the initial conditions 
\begin{eqnarray} 
p_{\rm sec}(\phi=0) &=& p^*, 
\label{4.14} \\ 
e_{\rm sec}(\phi=0) &=& e^* 
  \biggl[ 1 + \epsilon_{\rm c} \frac{4 + e^*}{4 e^*} \biggr], 
\label{4.15} \\ 
\omega_{\rm sec}(\phi=0) &=& 
-\epsilon_{\rm rr} \frac{8 + 5 e^* + 4 e^{*2}}{4 e^*}, 
\label{4.16} \\ 
t_{\rm sec}(\phi=0) &=& \frac{1}{4} \epsilon_{\rm rr} 
\sqrt{\frac{p^{*3}}{M}} \biggl[ \frac{3+4e^*}{(1+e^*)^2} 
\nonumber \\ & & \qquad \qquad \quad \mbox{} 
- \frac{3-4 e^{*2}}{(1-e^{*2})^{3/2}} \biggr] \quad
\label{4.17}
\end{eqnarray} 
in order to reproduce precisely the secular evolution predicted by the 
multi-scale approximation. 

The differential equations for $p_{\rm sec}$, $e_{\rm sec}$, and 
$\omega_{\rm sec}$ are easy to motivate:
Eqs.~(\ref{4.10})--(\ref{4.12}) are the same as Eq.~(\ref{3.19}), and
they can be obtained directly by submitting the exact equations
(\ref{3.9})--(\ref{3.11}) to an averaging procedure. The differential
equation for $t_{\rm sec}$, however, is not so easy to justify. It is 
not reproduced by averaging Eq.~(\ref{3.12}) over $\phi$, which would
fail to account for the important conservative correction proportional 
to $\epsilon_{\rm c}$; the averaging would also remove the Keplerian
oscillations of the time function. And the initial 
values of Eqs.~(\ref{4.14})--(\ref{4.17}) cannot be justified at all
without knowledge of the oscillatory terms in the multi-scale
approximation. To integrate the differential equations
with the approximate initial conditions $p_{\rm sec}(0) = p^*$, 
$e_{\rm sec}(0) = e^*$, $\omega_{\rm sec}(0) = 0$, and 
$t_{\rm sec}(0) = 0$ would produce solutions that are offset from the
exact solutions by quantities of order $\epsilon$. It is noteworthy
that while the corrections to $p_{\rm sec}(0)$, $\omega_{\rm sec}(0)$,
and $t_{\rm sec}(0)$ come as additive terms that become increasingly 
irrelevant as $\phi$ increases, the correction to $e_{\rm sec}(0)$
comes as a {\it multiplicative factor}; this correction never becomes 
irrelevant. 

Our main conclusion is this: The secular approximation defined by
the system of equations (\ref{4.10})--(\ref{4.17}) would be very
difficult to formulate without prior knowledge of the exact solution,
as represented by the faithful multi-scale approximation. This
conclusion reflects the lesson learned from the illustrative example
described in Sec.~II.  

\subsection{Secular approximation: $t$-average} 

The secular approximation considered in the preceding subsection is
obtained by removing the oscillations in $\phi$ from the exact
expressions for the orbital elements. Because we are ultimately
interested in the time behavior of the elements, it is perhaps more
meaningful to define the secular approximation by averaging with
respect to $t$ instead of $\phi$. In this alternative secular
approximation, we write   
\begin{eqnarray}  
p_{\rm sec}(\phi) = \langle p \rangle_t(\phi), \qquad
e_{\rm sec}(\phi) = \langle e \rangle_t(\phi),  
\nonumber \\ 
\omega_{\rm sec}(\phi) = \langle \omega \rangle_t(\phi), \qquad
t_{\rm sec}(\phi) = \langle t \rangle_t(\phi)  
\label{4.18}
\end{eqnarray}
in place of Eq.~(\ref{4.9}), where the time average is defined as in
Eq.~(\ref{2.9}), 
\begin{equation}
\langle q \rangle_t := 
\frac{\int_{\phi-\pi}^{\phi+\pi} q(\phi') (dt/d\phi')\,
  d\phi'}{\int_{\phi-\pi}^{\phi+\pi} (dt/d\phi')\, d\phi'}. 
\label{4.19}
\end{equation} 
The oscillations contained in $dt/d\phi$, as revealed in
Eq.~(\ref{3.32}), ensure that this version of the secular
approximation is quite distinct from the version examined in the
preceding subsection. 

Performing the calculations produces 
\begin{eqnarray} 
p_{\rm sec} &=& p_0,
\label{4.20} \\ 
e_{\rm sec} &=& e_0 + \frac{1}{4} (\epsilon_{\rm c} p^*) 
\frac{2 + e_0^2 - 2(1-e_0^2)^{3/2}}{e_0 p_0}, 
\label{4.21} \\ 
\omega_{\rm sec} &=& \omega_0 - \frac{1}{4} (\epsilon_{\rm rr}
p^{*3/2}) \frac{2 + 5e_0^2 - 2(1-e_0^2)^{3/2}}{e_0^2p_0^{3/2}}. 
\qquad  
\label{4.22}
\end{eqnarray} 
It is clear that these expressions do not agree with those of the
preceding subsection, in which we made the assignments 
$p_{\rm sec} = p_0$, $e_{\rm sec} = e_0$, and $\omega_{\rm sec} =
\omega_0$. 

In view of the complexity involved, we shall not attempt to find an
explicit expression for $t_{\rm sec}$. Nor shall we prolong the
discussion by writing down dynamical equations and initial conditions
for $p_{\rm sec}$, $e_{\rm sec}$, $\omega_{\rm sec}$, and 
$t_{\rm sec}$. We can simply jump to the main conclusion, which is the
same as in the preceding subsection: The dynamical equations and
initial conditions associated with this version of the secular
approximation would be very difficult to formulate without prior 
knowledge of the exact solution. Once more this conclusion reflects
the lesson learned from the illustrative example described in Sec.~II.   

\section{Conclusions}  

We examined the motion of a charged particle in a weak gravitational
field. In addition to the Newtonian gravity $\bm{g}$ exerted by a
large body of mass $M$, the particle is subjected to the
electromagnetic self-force described by Eq.~(\ref{1.2}). As we have
argued in Sec.~I, this toy problem shares many of the features of the
gravitational self-force problem, and yet it is sufficiently simple
that it can be solved completely with simple numerical methods, and
virtually completely with simple analytical methods. 

After subjecting the equations of motion to a multi-scale analysis in
Sec.~III, we summarized our main results in Sec.~IV and investigated
the main themes introduced in Sec.~I. We first examined the roles of 
the conservative and radiation-reaction pieces of the self-force. We
showed that the radiation-reaction force drives secular changes in the 
principal orbital elements $p$ and $e$, while the conservative force
drives secular changes in the positional element $\omega$ as well as
in the time function $t(\phi)$. This led us to our first conclusion: 
\begin{verse} 
The radiative approximation to the true self-force does not account
for the secular changes in all the orbital elements; this gives rise 
to an important phase mismatch between an orbital evolution driven
by the radiation-reaction force, and one driven by the true
self-force. The radiative approximation does not achieve the goals of
a secular approximation. 
\end{verse} 
This was also the conclusion of our previous work (paper I: 
Ref.~\cite{pound-poisson-nickel:05}), but we believe that we have 
established these statements more firmly in this work. In addition,
the source of the phase mismatch was correctly identified here, 
while it was attributed incorrectly in paper I: it is the conservative  
correction in the time function that is mostly responsible for the 
dephasing, and not the secular change in $\omega$.   

We next considered the issue of formulating secular approximations
to the dynamical equations that govern the evolution of the orbital 
elements. Having access to a faithful, analytical representation of
this evolution, as provided by the multi-scale approximation, it was
an easy task to perform averages and to obtain expressions that
capture the secular changes in the orbital elements (and the time
function). And having access to those expressions, it was again an
easy task to identify the differential equations that govern their 
behavior, as well as the appropriate initial conditions. The issue, of
course, is whether the simplified dynamical equations, those that
would govern the purely secular changes in the orbital elements, could
be obtained directly in a context in which the exact solutions are not
known. Our answer is in the negative, and this led us to our second
conclusion: 
\begin{verse} 
A secular approximation to the exact differential equations and 
initial conditions, designed to capture the secular changes in the
orbital elements and to discard the oscillations, would be very
difficult to formulate without prior knowledge of the exact
solution. While some of the approximate differential equations can be
obtained by submitting the exact equations to an averaging procedure,
other equations cannot be obtained so simply. And even if the correct 
differential equations can be identified, their integration must
proceed from initial conditions that differ from the exact initial
conditions; the difference is determined by the oscillations, and
those must be known before the approximate initial conditions can be
prescribed. 
\end{verse} 
In addition to these issues, the formulation of a secular
approximation must resolve a fundamental ambiguity: Which oscillations
are to be removed? In our analysis we had to distinguish carefully
between taking a $\phi$-average to remove oscillations in $\phi$, and
taking a $t$-average to remove oscillations in $t$. Different choices
lead to different secular approximations, different dynamical
equations, and a different prescription for initial conditions. 

The work presented here leaves a number of questions to be
examined. The most important one is this: Do the conclusions of this 
paper have any relevance to the gravitational self-force? While our 
analysis of the electromagnetic self-force leaves no room for
controversy, the question of how our results will transfer to the   
more interesting case of the gravitational self-force might be cause
for debate. We believe that the analogy between the electromagnetic 
and gravitational self-forces is close, we believe that our general 
conclusions do carry over to this case, and we believe that our work 
serves as a useful cautionary tale for the gravitational
self-force. But we admit that the analogy relies on the usual 
formulation of the gravitational self-force in the Lorenz gauge,  
and that the analogy may be lost in alternative formulations ---
the gravitational self-force is not gauge invariant, and its effect on
the description of orbital evolutions will depend on the choice of
gauge. For example, Mino \cite{mino:05a, mino:06} has proposed a
formulation of the gravitational self-force in a ``radiation-reaction
gauge'' in which the full self-force is equal (for a
radiation-reaction time) to the radiative self-force. In Mino's
proposed formulation, the radiative approximation is exact over a
radiation-reaction time, and the issues raised here may not at all be
relevant. How our conclusions might apply to the gravitational case is
indeed a controversial topic, but we consider its discussion to be
beyond the scope of this work. Indeed, this paper is concerned with
the electromagnetic self-force, and the case of the gravitational
self-force is considered separately in a companion paper
\cite{pound-poisson:07a}. In our companion work we argue that the
Lorenz-gauge formulation of the gravitational self-force is physically
meaningful, that the Lorenz gauge is most likely to keep quantities
other than the self-force (such as the gravitational potentials) under
control, and that the conclusions of this paper do carry over to the
gravitational case.      

\begin{acknowledgments} 
This work was supported by the Natural Sciences and Engineering
Research Council of Canada.     
\end{acknowledgments} 

\appendix 
\section{Keplerian motion} 

In this Appendix we provide a complete description of Kepler's
problem. This material is well known, and can be found in any textbook
on celestial mechanics (see, for example,
Ref.~\cite{murray-dermott:99}), but we include it here for
completeness and as a way of defining our notation.   

Two bodies of masses $m_1$ and $m_2$ move under their mutual
gravitational attraction. The equation of motion for the relative
position $\bm{r} := \bm{r}_1 - \bm{r}_2$ is 
\begin{equation} 
\bm{a} = \bm{g}, 
\label{A.1} 
\end{equation} 
where $\bm{a} := d^2 \bm{r}/dt^2$ is the relative acceleration vector, 
and $\bm{g} = -M \bm{r}/r^3$ is the gravitational field. Here $M = m_1
+ m_2$ is the total mass, and $r = |\bm{r}|$ is the distance between
the two bodies. We set $G=1$.  

Conservation of angular momentum implies that the motion takes place
within a fixed plane. We use polar coordinates $(r,\phi)$ in this
plane, and we resolve all vectors in the associated basis $(\unit{r}, 
\unit{\phi})$. The relation with the Cartesian description is $x =
r\cos\phi$, $y = r\sin\phi$, $\unit{r} = \cos\phi\, \unit{x} 
+ \sin\phi\, \unit{y}$, and $\unit{\phi} = -\sin\phi\, \unit{x} 
+ \cos\phi\, \unit{y}$. The position vector is $\bm{r} = r \unit{r}$,
the velocity vector is $\bm{v} = \dot{r}\, \unit{r} + r \dot{\phi}\,
\unit{\phi}$, and the acceleration vector is 
\begin{equation} 
\bm{a} = (\ddot{r} - r \dot{\phi}^2) \unit{r} 
+ \frac{1}{r} \frac{d}{dt} ( r^2 \dot{\phi} ) \unit{\phi}. 
\label{A.2}
\end{equation} 
An overdot indicates differentiation with respect to $t$. 

Equations (\ref{A.1}) and (\ref{A.2}) imply 
\begin{equation} 
r^2 \dot{\phi} = \mbox{constant} =: \sqrt{M p}, 
\label{A.3}
\end{equation} 
which defines the semilatus rectum $p$. We also have 
\begin{equation} 
\ddot{r} + \frac{M}{r^2} - \frac{Mp}{r^3} = 0, 
\label{A.4}
\end{equation}
which integrates to 
\begin{equation} 
\frac{1}{2} \dot{r}^2 - \frac{M}{r} + \frac{Mp}{2 r^2} 
= \mbox{constant} =: -\frac{M}{2p}(1-e^2). 
\label{A.5}
\end{equation} 
The constant is the system's conserved energy per unit reduced-mass,
and the last equation defines the eccentricity $e$.   

Eliminating time from Eqs.~(\ref{A.3}) and (\ref{A.4}) produces a
differential equation for $r(\phi)$ which integrates to 
\begin{equation} 
r(\phi) = \frac{p}{1 + e\cos(\phi - \omega)}, 
\label{A.6}
\end{equation} 
where $\omega$ is an additional constant of the motion. This equation
describes an off-centered ellipse of semi-major axis 
\begin{equation}
a = \frac{p}{1 - e^2} 
\label{A.7}
\end{equation} 
and eccentricity $e$. The constant $\omega$, known as {\it longitude
of periapsis}, determines the orientation of the ellipse in the
plane. The orbit is at periapsis $r = p/(1+e)$ whenever
$\cos(\phi-\omega) = 1$, and is at apoapsis $r = p/(1-e)$ whenever
$\cos(\phi-\omega) = -1$.  

Equations (\ref{A.3}) and (\ref{A.6}) imply 
\begin{equation}
\dot{r} = e \sqrt{\frac{M}{p}} \sin(\phi-\omega) 
\label{A.8}
\end{equation} 
and 
\begin{equation} 
\dot{\phi} = \sqrt{\frac{M}{p^3}} 
\bigl[ 1 + e\cos(\phi-\omega) \bigr]^2. 
\label{A.9}
\end{equation} 
This last equation integrates to 
\begin{equation} 
t(\phi) = t_{\rm peri} + \sqrt{\frac{p^3}{M}} 
\int_\omega^\phi \frac{d\phi'}{\bigl[ 
1 + e\cos(\phi'-\omega) \bigr]^2}  
\label{A.10}
\end{equation} 
and determines the time. The fourth (and final) constant of
integration $t_{\rm peri}$ is {\it time at periapsis}, and is such
that $t(\phi=\omega) = t_{\rm peri}$. According to Eq.~(\ref{A.10})
the orbital period is  
\begin{equation} 
P = \frac{2\pi}{n}, \qquad 
n := \sqrt{\frac{M}{a^3}}, 
\label{A.11}
\end{equation}
where $n$ is known as the {\it mean motion}.  

\section{Osculating orbital elements}  

In this Appendix we develop a method of osculating orbital elements
for the integration of the equations of motion associated with a
perturbed Keplerian orbit. The general idea is very old, and many
variations of this method can be found in the literature (see, for
example, Ref.~\cite{murray-dermott:99}). But we find that the version
presented here is perhaps a little unusual, while being especially
convenient and well suited to our purposes. For these reasons we judge
it worthwhile to develop it in full here.    

We consider the equations of motion 
\begin{equation} 
\bm{a} = \bm{g} + \bm{f}, 
\label{B.1}
\end{equation} 
in which $\bm{f}$ is a perturbing force (divided by the system's
reduced mass) that depends on the relative position vector $\bm{r}$
and (possibly) the relative velocity vector $\bm{v}$. (The notation is
introduced in Appendix A.) We seek to integrate Eq.~(\ref{B.1}) for
$\bm{r}(t)$ using a {\it method of osculating orbital elements}. We
assume, for simplicity, that the perturbing force can be decomposed as   
\begin{equation} 
\bm{f} = R\, \unit{r} + S\, \unit{\phi}, 
\label{B.2}
\end{equation} 
so that it lies within the orbital plane. The perturbed orbit,
therefore, will stay within the same plane.    

\subsection{First formulation} 

Let 
\begin{equation} 
I^A := \{ p, e, \omega, t_{\rm peri} \} 
\label{B.3}
\end{equation}
collectively stand for the Keplerian orbital elements introduced in
Appendix A, let  
\begin{equation} 
\bm{r}_{\rm K}(I^A, t) 
\label{B.4} 
\end{equation} 
stand for the position vector of a Keplerian orbit, and let 
\begin{equation} 
\bm{v}_{\rm K}(I^A, t) 
\label{B.5} 
\end{equation} 
be the Keplerian velocity vector. The method of osculating elements 
states that the perturbed motion is described at all times by
Eqs.~(\ref{B.4}) and (\ref{B.5}), but that the orbital elements
acquire a time dependence. In mathematical terms, the position vector
of the perturbed orbit is 
\begin{equation} 
\bm{r} = \bm{r}_{\rm K}\bigl(I^A(t),t\bigr)
\label{B.6} 
\end{equation} 
and its velocity vector is 
\begin{equation} 
\bm{v} = \bm{v}_{\rm K}\bigl(I^A(t),t\bigr). 
\label{B.7} 
\end{equation} 

Differentiating Eq.~(\ref{B.6}) with respect to time yields 
\[
\bm{v} = \frac{\partial \bm{r}_{\rm K}}{\partial I^A} 
\frac{d I^A}{dt} + \frac{\partial \bm{r}_{\rm K}}{\partial t}. 
\]
The second term, in which $\bm{r}_{\rm K}$ is differentiated while
keeping $I^A$ constant, is recognized as $\bm{v}_{\rm K}$, the
Keplerian velocity vector. Comparing with Eq.~(\ref{B.7}) gives 
\begin{equation} 
\frac{\partial \bm{r}_{\rm K}}{\partial I^A} \dot{I}^A = 0. 
\label{B.8}
\end{equation}
Differentiating Eq.~(\ref{B.7}) with respect to time yields 
\[
\bm{a} = \frac{\partial \bm{v}_{\rm K}}{\partial I^A} 
\frac{d I^A}{dt} + \frac{\partial \bm{v}_{\rm K}}{\partial t}. 
\]
The second term gives $\bm{g}$, and comparing with Eq.~(\ref{B.1})
gives 
\begin{equation} 
\frac{\partial \bm{v}_{\rm K}}{\partial I^A} \dot{I}^A = \bm{f}. 
\label{B.9}
\end{equation}
Equations (\ref{B.8}) and (\ref{B.9}) can be solved for $\dot{I}^A$ in
terms of the perturbing force. The equations of motion have become a
system of first-order differential equations for the orbital
elements. The method of osculating orbital elements therefore
transforms the original phase space spanned by $(\bm{r},\bm{v})$ into
a new phase space spanned by the coordinates $I^A$. In the planar
context considered here, the original phase space is spanned by $(r,
\phi, \dot{r}, \dot{\phi})$ while the new phase space is spanned by 
$(p, e, \omega, t_{\rm peri})$. 

Concretely the equations of motion are 
\begin{equation} 
\ddot{r} - r\dot{\phi}^2 + \frac{M}{r^2} = R, \qquad 
\frac{d}{dt} (r^2 \dot{\phi}) = r S. 
\label{B.10} 
\end{equation} 
By virtue of Eq.~(\ref{A.3}) and the osculating conditions of
Eqs.~(\ref{B.6}) and (\ref{B.7}), $r^2 \dot{\phi} = \sqrt{M p}$ and
the second of Eqs.~(\ref{B.10}) implies $r S = \frac{1}{2}
\sqrt{M/p}\, \dot{p}$. Inserting Eq.~(\ref{A.6}) yields 
\begin{equation} 
\dot{p} = 2 \sqrt{\frac{p^3}{M}} \frac{1}{1 + e\cos(\phi - \omega)} 
S,  
\label{B.11}
\end{equation} 
the new equation of motion for $p(t)$. 

To work out the remaining equations we substitute Eq.~(\ref{A.8}) into
the first of Eqs.~(\ref{B.10}). This gives  
\begin{eqnarray} 
R &=& -\dot{p}\, \frac{e}{2} \sqrt{\frac{M}{p^3}} \sin(\phi-\omega)  
+ \dot{e}\, \sqrt{\frac{M}{p}} \sin(\phi-\omega) 
\nonumber \\ & & \mbox{}
- \dot{\omega}\, e \sqrt{\frac{M}{p}} \cos(\phi-\omega), 
\label{B.12}
\end{eqnarray} 
after canceling out all Keplerian terms. An additional equation is
obtained by differentiating Eq.~(\ref{A.6}) with respect to time and
demanding that the result be compatible with Eq.~(\ref{A.8}). After
some algebra we obtain 
\begin{equation} 
0 = \dot{p} 
- \frac{p \cos(\phi-\omega)}{1 + e \cos(\phi-\omega)} \dot{e}  
- \frac{e p \sin(\phi-\omega)}{1 + e \cos(\phi-\omega)} \dot{\omega}. 
\label{B.13}
\end{equation} 

Equations (\ref{B.11})--(\ref{B.13}) imply 
\begin{eqnarray} 
\dot{e} &=& \sqrt{\frac{p}{M}} \biggl[ \sin(\phi-\omega)\, R 
\nonumber \\ & & \mbox{}
+ \frac{e + 2 \cos(\phi-\omega) + e\cos^2(\phi-\omega)}
       {1 + e\cos(\phi-\omega)}\, S \biggr] 
\label{B.14} 
\end{eqnarray} 
and 
\begin{eqnarray} 
e\dot{\omega} &=& \sqrt{\frac{p}{M}} \biggl[ -\cos(\phi-\omega)\, R  
\nonumber \\ & & \mbox{}
+ \frac{ \sin(\phi-\omega) [2 +  e \cos(\phi-\omega)]}
       {1 + e\cos(\phi-\omega)}\, S \biggr].  
\label{B.15} 
\end{eqnarray} 
In these equations, $\phi$ is a function of time that must be obtained
by integrating Eq.~(\ref{A.9}),
\[
\dot{\phi} = \sqrt{\frac{M}{p^3}} 
\bigl[ 1 + e\cos(\phi-\omega) \bigr]^2, 
\]
in which $p$, $e$, and $\omega$ are now time-varying orbital
elements. 

Our system of equations currently leaves out $t_{\rm peri}$, the
fourth orbital element. An equation for $\dot{t}_{\rm peri}$, however,
will not be required.  

\subsection{Second formulation} 

The preceding system of equations achieves a cleaner structure if we 
change the independent variable from $t$ to $\phi$ via
Eq.~(\ref{A.9}). Writing, for example, $p' := dp/d\phi 
= \dot{p}/\dot{\phi}$, we obtain 
\begin{eqnarray} 
p' &=& \frac{2p^3}{M} \frac{1}{(1 + e c)^3}\, S, 
\label{B.16} \\ 
e' &=& \frac{p^2}{M} \biggl[ \frac{s}{(1 + e c)^2}\, R 
+ \frac{e + 2c + ec^2}{(1+ec)^3}\, S \biggr], 
\label{B.17} \\ 
e\omega' &=& \frac{p^2}{M} \biggl[ -\frac{c}{(1 + e c)^2}\, R 
+ \frac{s(2 + ec)}{(1+ec)^3}\, S \biggr], 
\label{B.18} \\ 
t' &=& \sqrt{\frac{p^3}{M}} \frac{1}{(1+ec)^2}, 
\label{B.19}
\end{eqnarray} 
where 
\begin{equation}
c := \cos(\phi-\omega), \qquad 
s := \sin(\phi-\omega). 
\label{B.20}
\end{equation} 
The first three equations for $p(\phi)$, $e(\phi)$, and $\omega(\phi)$
constitute a closed system that can be solved independently of
the fourth equation, which determines $t(\phi)$. These equations are
exact, they are convenient to deal with, and they can easily be
implemented numerically. (The equations are ill-behaved when $e \to
0$; a transformation to new variables $\alpha = e\cos\omega$, $\beta =
e\sin\omega$ eliminates this pathology.) It is understood that the
system of equations is accompanied by the Keplerian representation of
the motion, that is, equations such as 
$r(\phi) = p/[1 + e \cos(\phi-\omega)]$ and 
$r' = e p \sin(\phi-\omega) / [1 + e \cos(\phi-\omega)]^2$.   

The second formulation of the method can be understood as follows. Let 
\begin{equation} 
I^A := \{ p, e, \omega \} 
\label{B.21}
\end{equation}
collectively stand for the relevant orbital elements, let  
\begin{equation} 
\bm{r}_{\rm K}(I^A, \phi), \qquad t_{\rm K}(I^A, \phi)  
\label{B.22} 
\end{equation} 
stand for the position vector of a Keplerian orbit, parameterized by
longitude $\phi$, and let  
\begin{equation} 
\bm{r}'_{\rm K}(I^A, \phi) := \frac{\partial \bm{r}_{\rm K}}
  {\partial \phi}, \qquad 
t'_{\rm K}(I^A, \phi) := \frac{\partial t_{\rm K}}
  {\partial \phi}. 
\label{B.23} 
\end{equation} 
The Keplerian velocity vector can then be expressed as $\bm{v}_{\rm K} 
= \bm{r}'_{\rm K}/t'_{\rm K}$. 

The method of osculating elements states that the perturbed motion
continues to be described by Eqs.~(\ref{B.22}) and (\ref{B.23}), but
that the orbital elements acquire a $\phi$-dependence. In mathematical
terms, the position vector of the perturbed orbit is 
\begin{equation} 
\bm{r} = \bm{r}_{\rm K}\bigl(I^A(\phi),\phi\bigr), \qquad 
t = t_{\rm K}\bigl(I^A(\phi),\phi\bigr)
\label{B.24} 
\end{equation} 
and we impose also 
\begin{equation} 
\bm{r}' = \bm{r}'_{\rm K}\bigl(I^A(\phi),\phi\bigr), \qquad 
t' = t'_{\rm K}\bigl(I^A(\phi),\phi\bigr). 
\label{B.25} 
\end{equation} 
The first two equations are equivalent to Eq.~(\ref{B.6}), and the
last two equations are equivalent to Eq.~(\ref{B.7}). The second of 
Eqs.~(\ref{B.25}) is the same as Eq.~(\ref{B.19}). 

Differentiating Eq.~(\ref{B.24}) with respect to $\phi$ yields 
\[
\bm{r}' = \frac{\partial \bm{r}_{\rm K}}{\partial I^A} 
\frac{d I^A}{d\phi} + \frac{\partial \bm{r}_{\rm K}}{\partial \phi}.  
\]
Comparing with Eq.~(\ref{B.25}) gives 
\begin{equation} 
\frac{\partial \bm{r}_{\rm K}}{\partial I^A} I'_A = 0. 
\label{B.26}
\end{equation}
Differentiating Eq.~(\ref{B.7}) with respect to $\phi$ and dividing by 
$t'$ from Eq.~(\ref{B.25}) yields 
\[
\bm{a} = \frac{\bm{v}'}{t'} = \frac{1}{t'_{\rm K}} \biggl[ 
\frac{\partial \bm{v}_{\rm K}}{\partial I^A} 
\frac{d I^A}{d\phi} + \frac{\partial \bm{v}_{\rm K}}{\partial \phi}
\biggr].  
\]
The second term gives $\bm{g}$, and comparing with Eq.~(\ref{B.1}) 
gives 
\begin{equation} 
\frac{1}{t'_{\rm K}} \frac{\partial \bm{v}_{\rm K}}{\partial I^A} 
I'_A = \bm{f}.  
\label{B.27}
\end{equation}
Equations (\ref{B.26}) and (\ref{B.27}) can be solved for $I'_A$ in 
terms of the perturbing force, and the end result is the system of
Eqs.~(\ref{B.16})--(\ref{B.18}). In this formulation the method of
osculating orbital elements transforms the original phase space
spanned by $r(t)$, $\phi(t)$, $\dot{r}(t)$,  and $\dot{\phi}(t)$ into
a new phase space spanned by $p(\phi)$, $e(\phi)$, $\omega(\phi)$, and
$t(\phi)$.  

The second formulation of the method of osculating elements is
distinguished by the facts that it involves $\phi$ as a running
orbital parameter, and it removes $t_{\rm peri}$ from the list of 
phase-space variables. This formulation leads to the important
advantages that Eqs.~(\ref{B.16})--(\ref{B.18}) form a closed set 
of equations; these equations can be integrated first, and $t(\phi)$
can be recovered at a later stage by solving Eq.~(\ref{B.19}).  

\bibliography{../bib/master}
\end{document}